\documentclass[aps,prb, bibliography, twocolumn, reprint, showpacs, footnoteinbib, superscriptaddress]{revtex4-1}
\usepackage{graphicx}
\usepackage{amssymb}   
\usepackage{amsfonts}
\usepackage{amsmath}
\usepackage{dcolumn}   
\usepackage{bm}        
\usepackage[mathscr]{eucal}
\usepackage[dvipsnames]{xcolor}
\usepackage[colorlinks, linkcolor=OrangeRed,citecolor=RoyalBlue,urlcolor=NavyBlue]{hyperref}
\usepackage[all]{hypcap} 
\usepackage{mathtools}
\usepackage{wasysym}
\usepackage{graphicx}
\usepackage{tikz}
\usepackage[utf8]{inputenc}
\usepackage{calc}
\usepackage{accents}
\newcommand{\dbtilde}[1]{\accentset{\approx}{#1}}
\usepackage{dsfont}
\newcommand{\mc}{\mathcal}

\newcommand{\dg}{\dagger}
\newcommand{\sq}{\square}
\newcommand{\ra}{\rangle}
\newcommand{\la}{\langle}

\newcommand*{\citen}[1]{%
  \begingroup
    \romannumeral-`\x % remove space at the beginning of \setcitestyle
    \setcitestyle{numbers}%
    \cite{#1}%
  \endgroup   
}

\definecolor{purplepaper}{rgb}{0.34, 0.12, .50} 
\definecolor{yellowpaper}{RGB}{252, 170, 62}
\begin{document}

\title{Variational classical networks for dynamics in interacting quantum matter}
%Variationally  trained classical networks for dynamics in interacting quantum matter

\author{Roberto Verdel}
\affiliation{Max Planck Institute for the Physics of Complex Systems, N{\"o}thnitzer Stra{\ss}e 38, Dresden 01187, Germany}

\author{Markus Schmitt}
\affiliation{University of California, Berkeley, California 94720, USA}

\author{Yi-Ping Huang}
\affiliation{The Paul Scherrer Institute, Forschungsstrasse 111, 5232 Villigen, Switzerland}
\affiliation{Department of Physics, National Tsing Hua
University, Hsinchu 30013, Taiwan}

\author{Petr Karpov}
%\email{karpov@pks.mpg.de}
\affiliation{Max Planck Institute for the Physics of Complex Systems, N{\"o}thnitzer Stra{\ss}e 38, Dresden 01187, Germany}
\affiliation{National University of Science and Technology ``MISiS'', Moscow 119991, Russia}

\author{Markus Heyl}
\affiliation{Max Planck Institute for the Physics of Complex Systems, N{\"o}thnitzer Stra{\ss}e 38, Dresden 01187, Germany}

\date{\today}

\begin{abstract}
Dynamics in correlated quantum matter is a hard problem, as its exact solution generally involves a computational effort that grows exponentially with the number of constituents. While remarkable progress has been witnessed in recent years for one-dimensional systems, much less has been achieved for interacting quantum models in higher dimensions, since they incorporate an additional layer of complexity. In this work, we employ a variational method that allows for an efficient and controlled computation of the dynamics of quantum many-body systems in one and higher dimensions. The approach presented here introduces a variational class of  wave functions based on complex networks of classical spins akin to artificial neural networks,  which can be constructed in a controlled fashion. We provide a detailed prescription for such constructions and illustrate their performance by studying quantum quenches in one- and two-dimensional models. In particular, we investigate the nonequilibrium dynamics of a genuinely interacting  two-dimensional lattice gauge theory, the quantum link model, for which we have recently shown---employing the technique discussed thoroughly in this paper---that it features disorder-free localization dynamics [P. Karpov \textit{et al}., \href{https://doi.org/10.1103/PhysRevLett.126.130401}{Phys. Rev. Lett. {\bf 126}, 130401 (2021)}]. The present work not only supplies a framework to address purely theoretical questions but also could be used to provide a theoretical description of experiments in quantum simulators, which have recently seen an increased effort targeting two-dimensional geometries. Importantly, our method can be applied to any quantum many-body system with a well-defined classical limit.  
\end{abstract}

\maketitle
\section{Introduction} \label{Introduction}

One of the main challenges in quantum many-body dynamics is that, unless the model under study is exactly solvable, the numerical overhead required to find an exact solution grows, in general, exponentially with the number of degrees of freedom. In the last decades the development of powerful computational techniques has nevertheless seen impressive progress, largely motivated by the experimental advances in realizing and controlling isolated quantum systems far away from equilibrium~\cite{Greiner2002, Kinoshita2006, Bakr547, simon2011, Hild2014,PhysRevLett.114.083002, Labuhn2016, Choi1547, Bernien2017, Jurcevic2017, PhysRevX.7.041047, Lienhard2018, Erne2018,  Barredo2018, Guardado-Sanchez2018, PhysRevX.10.011042}.
The majority of the advances have been achieved for one-dimensional (1D) systems, for which there exist now a set of reliable methods that can simulate efficiently the dynamics of  lattice models. The primary example of such set of techniques is tensor network algorithms such as the time-dependent density matrix renormalization group (or its variants TEBD and tMPS)~\cite{Cazalilla2002,  PhysRevLett.91.147902, Vidal2004, Daley2004, White2004, Verstraete2004,  PhysRevLett.93.207205, PAECKEL2019167998}, which uses a matrix product state~\cite{Perez-Garcia2007, VerstraeteMPS, SCHOLLWOCK201196} representation of the wave function and solves the dynamics, for instance, via a Trotter decomposition of the evolution operator. Yet, this approach is generally restricted  due to a rapid growth of entanglement, and a substantial increment of its computational complexity in higher dimensions.  A recent alternative consists of encoding quantum states in various types of networks of classical degrees of freedom, such as artificial neural networks (ANNs)~\cite{Carleo602, Czischek2018, PhysRevLett.125.100503} and perturbative classical networks (pCNs)~\cite{Schmitt2018}. %The former approach gets the dynamics by employing a time-dependent variational principle (TDVP)~\cite{Kramer, carleo2012}, whereas the latter relies on a refined form of time-dependent perturbation theory
On the other hand, the description of quantum dynamics in higher dimensions is facing severe limitations. In spite of a few very recent efforts in two dimensions (2D) using tensor networks~\cite{Murg2007, PhysRevA.92.053629, Zaletel2015, hashizume2018dynamical, Czarnik2019, Hubig2019, PhysRevB.102.235132, hubig2019evaluation, PhysRevB.102.035115,10.21468/SciPostPhys.9.5.070}, artificial neural networks~\cite{Fabiani2019, PhysRevLett.125.100503, lpezgutirrez2019real}, or numerical linked cluster expansion\cite{10.21468/SciPostPhys.9.3.031}, solving the quantum dynamics of 2D (and higher-dimensional) interacting systems remains one of the central challenges in computational quantum physics.

In this work, we introduce a  numerical framework that allows for an efficient solution of the dynamics of quantum lattice models in one and higher dimensions. In our method, the many-body wave function is represented as a complex network of classical spins akin to ANNs,  with couplings among the spins  that are taken as variational parameters, and which are then optimized via a time-dependent variational principle (TDVP)~\cite{Kramer, Carleo602}, with the key advantage that the classical networks do not face numerical instabilities as they have been observed for ANNs~\cite{PhysRevLett.125.100503}. The  architecture  of these \emph{variational classical networks} (VCNs) can be derived systematically, in a  fashion similar to that in the case of their relative pCNs~\cite{Schmitt2018}. A crucial property of VCNs is that they inherit the \emph{controlled} character that arises from the perturbative nature of pCNs. In addition, the optimization step introduced with the TDVP allows us to mitigate several inherent drawbacks of pCNs, such as being forcibly limited to weak quantum fluctuations and short timescales. Furthermore, there exist situations, as discussed subsequently, where VCNs may bear a reduced  computational complexity compared to similar state-of-the-art techniques, while still yielding sufficiently accurate results.

We employ our technique to study the dynamics of  two quantum spin systems: First, we consider various quenches in the paradigmatic 1D transverse field Ising model  (TFIM), which is exactly solvable, thereby serving as testing ground for our method. Next, we tackle the more challenging case of a genuinely interacting 2D lattice gauge theory, namely, the quantum link model (QLM). As shown in our recent paper~\cite{PhysRevLett.126.130401}, this system features disorder-free localization~\cite{Smith2017, smithPRL119, Brenes2018, PhysRevB.102.165132, PhysRevB.102.224303}, a mechanism for ergodicity breaking in homogeneous systems due to local constraints imposed by gauge invariance. We find that the methodology presented here is particularly well suited for describing the dynamics of the 2D QLM in the nonergodic regime, allowing us to probe timescales and system sizes that could hardly be accessed with any other state-of-the-art computational scheme.

The present work serves as a companion paper to
Ref.~\citen{PhysRevLett.126.130401},
where we briefly introduced the VCN method but only  for the specific case of the 2D QLM and omitted some technical details.
Here we present the method in its full generality
and show how it can be applied to any interacting quantum system in arbitrary dimensions, as long as a classical limit can be defined for the model under consideration.

%\commentPRB{
%The present work serves as a companion paper to
%Ref.~\citen{2020arXiv200304901K},
%where we briefly introduced the VCN method but only  for the %specific case of the 2D QLM and omitted some technical details.
%Here we present the method in its full generality
%and show how it can be applied to any interacting quantum system in %arbitrary dimensions, as long as a classical limit can be defined for the model under consideration.}
%\textcolor{blue}{MS: Is this formulation too defensive?}

The outline of the remainder of this paper is as follows: In Sec.~\ref{method} we state the general problem and settings for the construction and subsequent application of our method.  Further, we show how to derive, in general, the structure of VCNs (Sec.~\ref{CNs}), and recall  the way in which the TDVP operates (Sec.~\ref{tdvp}).  Next, using our methodology we simulate quantum quenches in the paradigmatic 1D TFIM in Sec.~\ref{TFIM}, and in the genuinely interacting 2D QLM in Sec.~\ref{QLM}. Some concluding remarks, including possible applications to recent experiments in quantum simulators, are discussed in the last section.

\section{The method} \label{method}
 \begin{figure}[bt!]
	\centering
	\includegraphics[width=\columnwidth]{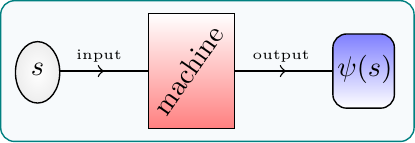}
	\caption{Schematic representation of a generative machine that takes as input a spin configuration $s$, and computes on-the-fly an approximation of the corresponding wave function amplitude $\psi(s)$. Such machine can be, for example, an ANN or, as employed in this work, a  variational classical network (VCN).}
	\label{fig:1}
\end{figure}

In this section, we explain the basic idea of our method. It contains two crucial ingredients: (\textit{i}) an efficient compression of the wave function in terms of networks of classical degrees of freedom, and (\textit{ii}) a suitable procedure to optimize such networks so as to get an accurate description of the encoded time-dependent quantum state.

\subsection{General problem and generative machines}\label{general}
Throughout this work, we shall consider systems of $N$ spin-1/2 degrees of freedom. We work in the computational basis of  spin configurations $s=(s_1, s_2, \dots, s_N)$, with $s_i=\uparrow, \downarrow$. In this representation, the state vector can be expanded as
\begin{equation}
\label{psi_vec}
|\psi(t)\ra= \sum_{\{s\}} \psi(s, t) |s\ra,
\end{equation}
where the amplitudes $\psi(s, t)\equiv \la s | \psi(t) \ra$ contain the full information about the system, from which, in principle, all physical quantities can be computed. However, there is a fundamental computational limitation associated with the expansion~\eqref{psi_vec}, as it involves  exponentially many terms.

A possible way to overcome this difficulty consists of using a generative machine that approximates the wave function \textit{on the fly}, rather than storing all the individual  coefficients; see Fig.~\ref{fig:1}. 
%This approach was first proposed with general-purpose ANNs as generative models~\cite{Carleo602}. 
This approach received increased attention recently, when general-purpose ANNs were proposed as generative models~\cite{Carleo602}. 
Here, we provide an alternative class of generative machines akin to ANNs, which can be constructed according to a controlled prescription as explained in the following. For the time being, however, let us consider a generic generative machine $\psi_{\eta}(s)$, such that
\begin{equation}
\label{machine}
\psi(s,t) \approx \psi_{\eta}(s),
\end{equation}
where $\eta$ refers to a set of complex parameters that carry the time dependence, i.e., $\eta \equiv \eta(t)=(\eta_1(t), \eta_2(t), \dots, \eta_K(t) )\in \mathbb{C}^K$, and which are optimized variationally.

A representation of the wave function in terms of such generative machine renders the problem of computing physical quantities tractable. In effect, the expectation value of an observable $O$ with matrix elements $\langle s|O|s'\rangle=O_{s,s'}$, can be written as
\begin{equation}
\label{obs}
\la \psi{_\eta} |O | \psi{_\eta} \ra= \sum_{\{s\}} |\psi_{\eta}(s)|^2O_{\eta}(s),
\end{equation}
where $O_{\eta}(s)=\sum_{\{s'\}}O_{s,s'} \psi_{\eta}(s')/\psi_{\eta}(s)$. Typical local observables are such that  $\la s|O|s'\ra$ is sparse. Consequently, getting $O_{\eta}(s)$  requires only a polynomial computational overhead. Hence, assuming a normalized wave function, the expectation value~\eqref{obs} can be calculated efficiently via a Monte Carlo sampling of the distribution $ |\psi_{\eta}(s)|^2$. Note that a compression of the wave function such as Eq.~\eqref{machine} will be efficient as long as the overall number $K$ of  parameters  is significantly less than the dimension of the Hilbert space. 

Let us also note that, in general, a generative machine refers to a model that can generate samples according to some target distribution, which in this case is $|\psi(s, t)|^2$. Remarkably, generative machines such as that in Eq.~\eqref{machine}, which we shall regard in the following, not only achieve the task mentioned above but also give direct access to the complex amplitudes $\psi(s, t)$.  
 
\subsection{Variational Classical Networks}\label{CNs}
As mentioned before, this work aims to introduce a class of adequate generative machines to represent the many-body wave function. In the following, we give a controlled prescription to construct such generative models.

\subsubsection{General settings}\label{settings}
Let us consider a Hamiltonian of the form 
\begin{equation}
\label{H}
H=H_0 + \gamma V,
\end{equation}
where $H_0$ represents a classical system in the sense that it is diagonal in the  computational basis,
\begin{equation}
\label{Hdiagonal}
H_0 |s\rangle =E_{s}|s\rangle,
\end{equation}
and the off-diagonal perturbation $\gamma V$, with $\gamma$  playing the role of a small parameter, accounts for quantum fluctuations.% that induce transitions between the eigenstates of $H_0$.

We are interested in the nonequilibrium dynamics generated by the Hamiltonian~\eqref{H}. This can be obtained by solving the time-dependent Schr\"odinger equation, which admits the formal solution (in units such that $\hbar =1$)
\begin{equation}
\label{sol}
|\psi(t)\ra= \mathrm{e}^{-iHt}|\psi_0\ra,
\end{equation}
where $|\psi_0\rangle$ denotes the initial state. In general, it is challenging to determine the action of the evolution operator $\mathrm{e}^{-iHt}$, onto the basis vectors. However, whenever the Hamiltonian $H$ can be split as in Eq.~\eqref{H},  it is possible to carry out a perturbative treatment by working in the interaction picture, in which the evolution operator can be written as
\begin{equation}
\label{inter_picture}
\mathrm{e}^{-iHt}=\mathrm{e}^{-iH_0t}W_{\gamma}(t),
\end{equation}
where
\begin{equation}
\label{inter_picture2}
W_{\gamma}(t)=\mc{T} \exp\Big[-i \gamma \int_0^t \mathrm{d}t' V(t') \Big],
\end{equation}
where $\mc{T}$ is the time-ordering operator, and with $V(t)$ satisfying the equation of motion:
\begin{equation}
\label{eom}
-i \frac{\mathrm{d}}{\mathrm{d}t}V(t)=[H_0,V(t)].
\end{equation}
Within these settings, the many-body wave function amplitudes are given by
\begin{equation}
\label{psi}
\psi(s,t) = \mathrm{e}^{-iE_{s}t}\la s |W_{\gamma}(t)| \psi_0\ra.
\end{equation}
The task now is to calculate the right-hand side Eq.~\eqref{psi}. Classical networks provide a  possible solution, as detailed below.

\subsubsection{Cumulant expansion and pCNs}\label{cumulantExp}
The right-hand side of Eq.~\eqref{psi} can be computed in a controlled way by means of a cumulant expansion~\cite{kubo62}, namely,
\begin{align}
\label{cumulant}
\la s |W_{\gamma}(t)| \psi_0\ra&= \la s|\psi_0 \ra  \exp \Bigg[\sum_{n=1}^\infty \frac{(-i\gamma)^n}{n!} \int_0^t \mathrm{d}t_1 \int_0^t \mathrm{d}t_2  \nonumber \\
&\cdots\int_0^t \mathrm{d}t_n \la \mc{T} V(t_1) V(t_2)\cdots V(t_n)\ra_c\Bigg],
\end{align}
where  $\la \cdot \ra_c $ denotes the cumulant average. For example, for the lowest-order corrections, we have
\begin{align}
\label{cumulant_averages}
 \la A\ra_c&\equiv\frac{\langle s|A|\psi_0\rangle}{\langle s|\psi_0\rangle}, \\
 \label{cumulant2}
 \la AB\ra_c&\equiv\frac{\langle s|AB|\psi_0\rangle}{\langle s|\psi_0\rangle} - \frac{\langle s|A| \psi_0\rangle \langle s |B|\psi_0 \rangle}{\langle s|\psi_0\rangle^2}.
 \end{align}
 This expansion allows us to write down the wave function as 
\begin{equation}
\label{pCN}
\psi(s,t) =  \mathrm{e}^{\mc{H}_\mathrm{eff}(s,t)},
\end{equation}
with $\mc{H}_\mathrm{eff}(s,t)$ defined by Eqs.~\eqref{psi} and \eqref{cumulant}. Representations of the wave function in the form of a Boltzmann-like factor are quite adequate to compute physical observables via a Monte Carlo procedure~\cite{carleo2012, carleoPRA, carleoPRX, Schmitt2018}. In order to gain some  insight about the physical content of the function $\mc{H}_\mathrm{eff}(s,t)$, let us restrict ourselves for a moment to a simple initial product state, namely, an equally weighted superposition of the spin configurations:
\begin{equation}
\label{x-initialState}
|\psi_0\ra  = \lvert\rightarrow \ra \equiv \bigotimes_{i=1}^N \frac{1}{\sqrt{2}}\big[|\uparrow\ra_i +|\downarrow\ra_i \big].
\end{equation}
This initial state  is particularly convenient as $\psi_0(s)= 2^{-N/2}$ for all $s$, and hence  $\psi_0(s)$ drops out in all the cumulant averages. In this scenario, and upon performing the integrals in the cumulant expansion \eqref{cumulant}, the  function $\mc{H}_\mathrm{eff}$ adopts, in general,  the following form
\begin{equation}
\label{Heff}
\mathcal{H}_\mathrm{eff}(s, t)=\sum_l C_l(t) \Phi_l(s).
\end{equation}
That is, $\mc{H}_\mathrm{eff}$ can be regarded as the effective Hamiltonian of a classical spin system with complex couplings $C_l(t)$, and with  spin interactions given by the functions $ \Phi_l(s)$, which are local provided that the quantum Hamiltonian is local, too. Situations where quenches from the initial state \eqref{x-initialState} are of physical interest are discussed in posterior sections.
 
In some cases, it is possible to recast the systems defined by Eq.~\eqref{Heff}, as conventional classical statistical mechanical models. For example, the effective model corresponding to the 1D TFIM, discussed later in detail (see Sec.~\ref{TFIM}), contains, up to first order in the cumulant expansion,  the following terms~\cite{Schmitt2018}:
\begin{equation}
\label{pCNIsing}
\Phi_1(s) = \sum_{i} s_i s_{i+1},  \hspace{0.4cm} \Phi_2(s) = \sum_i s_i s_{i+2},
\end{equation}
which define a 1D classical Ising model with nearest and next-to-nearest neighbor interactions.

\begin{figure}[bt!]
	\centering
	\includegraphics[width=1\columnwidth]{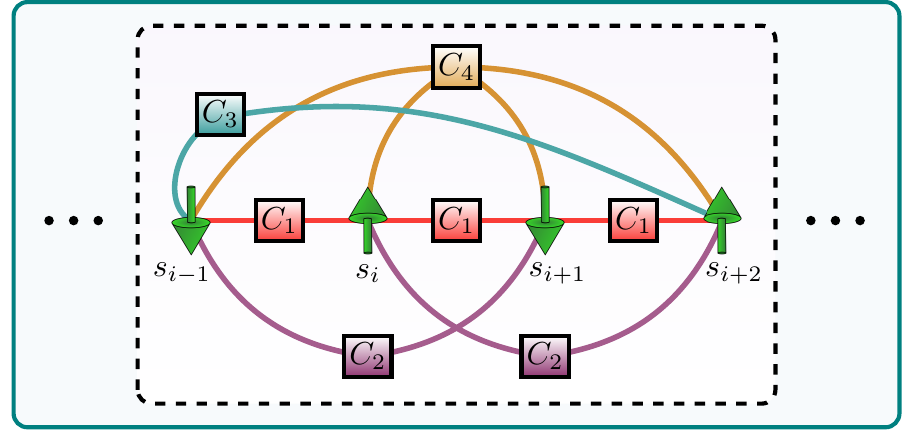}
	\caption{Local structure of a classical network for the 1D TFIM in a translationally invariant lattice, containing up to second-order terms. The network's connectivity is defined by functions such as those in Eq.~\eqref{pCNIsing} [see also Eqs.~\eqref{H1TFIM} and~\eqref{H2TFIM}, for this particular example]. The nodes of the network correspond to classical spins $s_i$, whereas the links of the network are specified by the couplings $C_l$.}
	\label{fig:network}
\end{figure}

On the other hand, the system \eqref{Heff} can also be visualized as a  \textit{network} of classical spins with  connectivity specified by the functions $ \Phi_l(s)$. Hence, the models defined by an effective Hamiltonian as in Eq.~\eqref{Heff} are called perturbative classical networks (pCNs)\cite{Schmitt2018}. In Fig.~\ref{fig:network}, we display the network representation, up to second order, of the classical spin model that emerges  when considering a translationally invariant 1D TFIM.

%with the lowest order corrections being specified by the previous formulas, that is, 
%\begin{equation}
%\label{H0}
%\mathcal{H}^{(0)}(s,t)=-iE_{s}t + \ln(\psi_0(s)),
%\end{equation}
%\begin{equation}
%\label{H1}
%\mathcal{H}^{(1)}(s,t)=-i \gamma \int_0^t \mathrm{d}t' \frac{\langles|V(t')|\psi_0\rangle}{\langle s|\psi_0\rangle},
%\end{equation}
%\begin{align}
%\label{H2}
%\mathcal{H}^{(2)}(s,t)=- \gamma^2 &\int_0^t \mathrm{d}t'\int_0^{t'} \mathrm{d}t'' \Bigg( \frac{\langle s|V(t')V(t'')|\psi_0\rangle}{\langle s|\psi_0\rangle} \nonumber \\ 
%	& - \frac{\langle s|V(t')| \psi_0\rangle \langle s |V(t'')|\psi_0 \rangle}{\langle s|\psi_0\rangle^2} \Bigg) .
%\end{align}

Let us emphasize that our approach works as well for initial states other than the equally weighted superposition \eqref{x-initialState}. Importantly, this is  true not only for translationally invariant initial states but also for nonuniform ones, as will be shown later. To consider a more general case, let us assume that the perturbation consists of a sum of local terms, namely, $V=\sum_\alpha v_\alpha$, where the overall number of terms is polynomial in system size. Thus, we can write an equation of motion for each of the individual terms: 
\begin{equation}
\label{eomv}
-i \frac{\mathrm{d}}{\mathrm{d}t}v_\alpha(t)=[H_0,v_\alpha(t)]=\mathrm{e}^{iH_0t} [H_0,v_\alpha]\mathrm{e}^{-iH_0t}.
\end{equation}
The commutator $[H_0, v_\alpha]$  measures, essentially, the energy difference between two eigenstates of $H_0$ in a transition induced by $v_\alpha$, i.e., $v_\alpha|s_1\ra= |s_2\ra$. Indeed, one can readily prove that 
\begin{equation}
\label{comm}
[H_0, v_\alpha]|s_1\ra = (E_{s_2}- E_{s_1}) v_\alpha|s_1\ra.
\end{equation}
Therefore, Eq.~\eqref{eomv} can be rewritten as  
\begin{equation}
\label{eom2}
\frac{\mathrm{d}}{\mathrm{d}t}v_\alpha(t)=i \Omega_\alpha v_\alpha(t),
\end{equation}
with $\Omega_\alpha$ being a diagonal operator in the computational basis, which measures the energy difference in a transition induced by $v_\alpha$. This equation admits the formal solution $v_\alpha(t)=\mathrm{e}^{i\Omega_\alpha t} v_\alpha$. Thus, if the effective Hamiltonian is written as $\mc{H}_\mathrm{eff}=\sum_{n=0}^\infty \mc{H}^{(n)}$, where the zeroth-order term is $\mc{H}^{(0)}(s, t)= -i E_{s} t + \ln(\psi_0(s))$,  and the subsequent orders are defined by Eq.~\eqref{cumulant}, one can write, for example, the first-order correction as
\begin{equation}
\label{H1}
\mc{H}^{(1)}(s,t)=-i \gamma \sum_{\{s'\}} \frac{\psi_0(s')}{\psi_0(s)} \sum_\alpha \la s|v_\alpha| s'\ra \int_0^t \mathrm{d}t' \mathrm{e}^{i \Omega_\alpha(s) t'},
\end{equation}
and likewise for higher-order terms. Note that the matrix $\la s|v_\alpha|s'\ra$ is typically sparse for physical systems with few-body couplings.

Let us remark that the cumulant expansion \eqref{cumulant} goes beyond conventional time-dependent perturbation theory, since the corrections considered here effectively account for a resummation of several terms that appear in a standard perturbative expansion~\cite{Schmitt2018}. However, pCNs face their own limitations, too. In particular, they are inherently restricted to weak quantum fluctuations (small $\gamma$)  to ensure that we can safely truncate the  expansion~\eqref{cumulant}. Besides, the description of the evolution of observables will eventually break down, since resonant processes may be present, giving rise to secular terms that limit a correct description to timescales of  order  $\mc{O}(1/\gamma)$~\cite{Schmitt2018}. Nonetheless, one can still benefit from the framework introduced here, while mitigating the drawbacks mentioned before. This is achieved by constructing adequate \textit{variational} wave functions with a network architecture that is inherited from a corresponding pCN, as argued in the following.

\subsubsection{Variational ansatz}\label{variational}
The main idea of this work was already outlined in the previous paragraph: building upon the structure of an underlying pCN, one can construct variational classical networks (VCNs) that are then used as generative machines to compute the dynamics. Importantly, the resulting VCNs will inherit the \emph{controlled} character of the cumulant expansion, in  that the accuracy of the approximation can be improved systematically by reducing the value of $\gamma$, or by taking into account higher-order cumulants.

First, let us consider classical networks of the form given in Eq.~\eqref{Heff}.  The corresponding VCN can be obtained simply by  letting the couplings $C_l$ be variational parameters, that is,
\begin{equation}
\label{VCN}
\mc{H}_\mathrm{VCN}(s; \eta(t))=\sum_l \eta_l(t) \Phi_l(s),
\end{equation}
where $\eta(t)$ denotes a set of complex variational parameters. 

In the more general case such as Eq.~\eqref{H1}, one can build the corresponding VCN by noting that $\Omega_\alpha(s)$ take a finite number of discrete values for any $s$. Thus one can simply introduce a variational parameter for each value of $\Omega_\alpha(s)$. To fix ideas let us consider the first-order correction given in Eq.~\eqref{H1}, and let us denote as $\Lambda_{\Omega_{\alpha}}$ the set of all possible values of  $\Omega_\alpha(s)$.  After rewriting the integral 
\begin{equation}
\label{integral}
 \int_0^t \mathrm{d}t' \mathrm{e}^{i \Omega_\alpha(s) t'} = \sum_{\Omega \in \Lambda_{\Omega_\alpha}} \delta_{\Omega_\alpha(s), \Omega} \int_0^t \mathrm{d}t' \mathrm{e}^{i \Omega t'},
\end{equation}
%Then we introduce a variational parameter for value of $\Omega_\alpha(s)$, that is,
%\begin{equation}
%\label{eta}
%\eta_{\Omega_\alpha(s)}(t):=\sum_{\omega=-4}^{4} \delta_{\omega_{\sq}(), \omega} \eta_{\omega}^{(1)}(t),
%\end{equation}
we can  introduce a set of variational parameters so that the corresponding first-order variational effective Hamiltonian reads
\begin{align}
\label{HVCN}
\mc{H}_\mathrm{VCN}^{(1)}(s; \eta(t))=&-i \gamma\sum_{\{s'\}} \frac{\psi_0(s')}{\psi_0(s)} \sum_\alpha \la s|v_\alpha| s'\ra \nonumber \\
& \times \sum_{\Omega \in \Lambda_{\Omega_\alpha}} \delta_{\Omega_\alpha(s), \Omega}\eta_{\Omega}^{(1)}(t).
\end{align}
and likewise for higher-order terms. In either case, the concomitant wave function amplitudes take the form 
\begin{equation}
\label{psiVCN}
\psi_\mathrm{VCN}(s; \eta(t))=\mathrm{e}^{\mc{H}_\mathrm{VCN}(s; \eta(t))}.
\end{equation}

 Let us point out that, the more higher-order corrections are included in the architecture of a classical network, the more quantum correlations can, in principle, be taken into account. Thus, the cumulant expansion~\eqref{cumulant} provides us with a controlled procedure to generate generative machines of the form given in Eq.~\eqref{psiVCN}, which allows for a systematic addition of terms that can potentially encode more and more nonlocal quantum correlations. Before turning to the applications, let us discuss the variational procedure that is used to optimize the resulting VCNs.

\subsection{Time-dependent variational principle}\label{tdvp}

The TDVP~\cite{Kramer, carleo2012, carleoPRA,  Carleo602} is a way of optimizing a time-dependent variational ansatz $\psi_{\eta}(s)$, where $\eta$ denotes a set of complex time-dependent variational parameters, i.e., $\eta(t)=(\eta_1(t), \eta_2(t), \dots, \eta_K(t))$.  Such trial wave function could be, e.g., a Jastrow ansatz~\cite{Jastrow,Blass2016}, an ANN~\cite{Carleo602}, or, as presented in this work, a VCN. In essence, the TDVP is a procedure that establishes an equivalence between the time-dependent Schr\"odinger equation and a system of first-order differential equations that govern the dynamics of the variational parameters, namely,
\begin{equation}
\label{tdvp_eq}
\sum_{k'}\mc{S}_{k,k'} \dot{\eta}_{k'} = -i F_{k},
\end{equation}
where the overdot denotes differentiation with respect to time and with the following definitions:
\begin{equation}
\label{cov}
\mc{S}_{k, k'}:=\langle O_k^{\ast} O_{k'}\rangle-\langle O_k^{\ast} \rangle\langle O_{k'}\rangle,
\end{equation}
 which is the so-called covariance matrix, and 
 \begin{equation}
 \label{forces}
F_k:=\langle E_\mathrm{loc}O_k^{\ast}\rangle - \langle E_\mathrm{loc}\rangle \langle O_k^{\ast}\rangle.
 \end{equation}
 These quantities are expressed in terms of the  local energy $E_\mathrm{loc}(s):=\frac{\langle s |H|\psi_{\eta}\rangle}{\langle s|\psi_{\eta} \rangle}$, and the variational derivatives,  $O_k(s):= \frac{\partial \ln \psi_{\eta}(s)}{\partial \eta_k}$.

In order to quantify the accuracy of this TDVP, let us introduce the Fubini-Study metric $\mathscr{D}_\mathrm{FS}$, which measures the distance between the exact evolution during a small  time interval  $\delta t$:  $\mathrm{ e}^{-i\delta t H}|\psi_{\eta}\rangle$, and the variational evolution $ |\psi_{\eta+\delta \dot{\eta}}\rangle$. Its definition is the following:
\begin{equation}
\label{fubini}
  \mathscr{D}_\mathrm{FS}(\varphi,\phi)^2:=\arccos \Bigg( \sqrt{\frac{\la \varphi|\phi\ra\la\phi |\varphi\ra}{\la \varphi|\varphi\ra\la\phi |\phi\ra}}\Bigg)^2.
\end{equation}
Thus, one can define a relative residual error  as~\cite{Carleo602,PhysRevLett.125.100503}
\begin{equation}
\label{error}
r^2(t):=\frac{\mathscr{D}_\mathrm{FS}(|\psi_{\eta+\delta \dot{\eta}}\rangle, \mathrm{ e}^{-i\delta t H}|\psi_{\eta}\rangle)^2}{\mathscr{D}_\mathrm{FS}(|\psi_{\eta}\rangle,\mathrm{ e}^{-i\delta t H}|\psi_{\eta}\rangle)^2}
\end{equation}
 which can be measured, too, by performing a Monte Carlo sampling of $|\psi_{\eta}(s)|^2$. In practice, we use a second-order expansion of~\eqref{fubini} to compute Eq.~\eqref{error}. Moreover, in the following, we shall consider the integrated residual error $R^2(t):=\int_0^t \mathrm{d}t' r^2(t') $. Note that Eq.~\eqref{tdvp_eq} can be derived by minimizing the numerator in Eq.~\eqref{error} with respect to $\dot{\eta}^{\ast}$~\cite{Carleo602,PhysRevLett.125.100503}.

\section{Quenches in the 1D TFIM} \label{TFIM}
\subsection{Model}

As a first illustration of our method, we study several quantum quenches in the archetypal 1D TFIM, whose Hamiltonian for $N$ spins on a ring reads

\begin{equation}
\label{Hising}
H_\mathrm{TFIM}= -J \sum_{ i=1}^N \sigma_i^z \sigma_{i+1}^z -h\sum_{i=1}^N \sigma_i^x, 
\end{equation}

\noindent where $\sigma_i^\mu$ ($\mu=x, y, z$) are the Pauli matrices at site $i$, $J > 0$ is the exchange constant that sets the overall energy scale, and $h$ is a transverse magnetic field. %Unless otherwise stated, periodic boundary conditions are considered. 

Let us recall some features of the model in Eq.~\eqref{Hising}. First of all, the 1D TFIM is integrable by means of a Jordan-Wigner transformation~\cite{PFEUTY197079}; hence, comparison with  analytical solutions is at our disposal.  Moreover, this model features both equilibrium~\cite{sachdev_2011} and dynamical~\cite{heyl2013} quantum phase transitions. Indeed, the Hamiltonian~\eqref{Hising} undergoes an equilibrium quantum phase transition at $h_c/J=1$~\cite{sachdev_2011}, where the critical point separates a ferromagnetic phase ($h< h_c$) from a paramagnetic one ($h>h_c$). Its dynamical quantum phase transition (DQPT) is signaled by non-analyticities in the many-body dynamics~\cite{heyl2013, Heyl_2018}, and occurs when quenching across the underlying equilibrium quantum phase transition. In this respect, it is interesting to study quenches that cross the critical point. Below, we shall consider quenches from the paramagnetic point $h_0=\infty$, which corresponds to the initial state in Eq.~\eqref{x-initialState}, to both the ferromagnetic and the paramagnetic phases (see details below). Note that the first type of quench is precisely of interest for the study of DQPTs. From an experimental viewpoint, both probing the dynamics of this model and  engineering the relevant initial state in Eq.~\eqref{x-initialState} are now feasible tasks with current technologies in quantum simulators in  various settings~\cite{simon2011, Bernien2017, Jurcevic2017}.

Before discussing the details of the quench dynamics simulations and the corresponding results, let us first construct the VCNs associated to the 1D TFIM.
 
\subsection{VCNs for the 1D TFIM}\label{CN_TFIM}

The corresponding pCNs for TFIMs have been recently derived elsewhere~\cite{Schmitt2018}. Here, we review the main steps of such calculations. First of all, according to the general settings established in Sec.~\ref{CNs}, we take the Ising interaction term as reference Hamiltonian, i.e., $H_0^\mathrm{TFIM}=-J\sum_{i=1}^N \sigma_i^z\sigma_j^z$, and the transverse field as the perturbation, namely, $\gamma V^\mathrm{TFIM}=-h \sum_i \sigma_i^x$ (identifying $\gamma =- h$).

Using the basic commutation relations of the Pauli matrices, one can readily show that 
\begin{equation}
\label{commIsing}
[H_0^\mathrm{TFIM}, \sigma_j^x]= -2 J (\sigma_{j-1}^z+\sigma^z_{j+1})\sigma_j^z\sigma_j^x,
\end{equation}
where  we emphasize again that this commutator measures the change in energy in a transition induced by $\sigma_j^x$, between eigenstates of $H_0^\mathrm{TFIM}$. The solution to the equation of motion Eq.~\eqref{eom2} for $\sigma_j^x(t)$ therefore reads
\begin{align}
\label{solSigmax}
\sigma_j^x(t) &=\mathrm{e}^{ -2 i (\sigma_{j-1}^z+\sigma^z_{j+1})\sigma_j^zJt}\sigma_i^x \nonumber \\
&= \Big[\cos^2(2Jt)-\sin^2(2Jt)\sigma_{j-1}^z\sigma^z_{j+1} \nonumber \\
  &\quad\quad-\frac{i}{2}\sin(4Jt)(\sigma_{j-1}^z+\sigma^z_{j+1})\sigma_j^z  \Big] \sigma_i^x, 
\end{align}
where  the second step follows from Euler's formula. This solution leads to a first-order pCN of the form anticipated in Eq.~\eqref{pCNIsing}. Indeed, plugging the solution~\eqref{solSigmax} in Eq.~\eqref{H1}  and using the fact that $\psi_0(s)= 2^{-N/2}$ for all $s$, for the initial state in Eq.~\eqref{x-initialState}, one gets
\begin{equation}
\label{H1TFIM}
\mc{H}^{(1)}_\mathrm{TFIM}=  C_0^{(1)}(t) N + C_1^{(1)}(t) \sum_{i=1}^N s_i s_{i+1} + C_2^{(1)}(t) \sum_{i=1}^N s_i s_{i+2}, 
\end{equation}
where the explicit form of the coefficients $C_l^{(1)}$ can be easily deduced from Eqs.~\eqref{solSigmax} and \eqref{H1}. As previously explained, the classical network defined above can be turned into a VCN, simply by regarding the couplings $C_l^{(1)}$ as variational parameters. 

\begin{figure}[t!]
	\centering
	\includegraphics[width=\columnwidth]{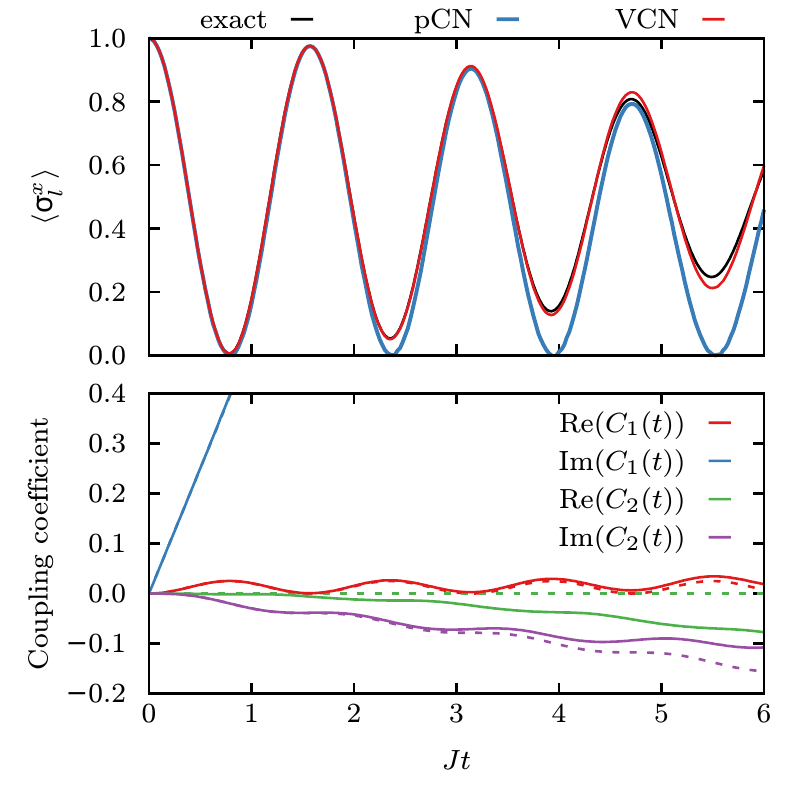}
	\caption{Comparison of first-order pCN, first-order VCN and the exact solution of the TFIM with  $N=50, h/J=0.1$. Top: Dynamics of the transverse magnetization $\la \sigma_l^x\ra$. The black curve shows the exact solution obtained via fermionization in the thermodynamic limit.  Bottom: Evolution of the perturbative (dashed lines) and variational (solid lines) couplings of the classical network defined in Eq.~\eqref{H1TFIM}. }
	\label{fig:tfim1}
\end{figure}
The perturbatively motivated structure of the VCN can be systematically expanded by straightforwardly plugging Eq.\ \eqref{solSigmax} into higher-order terms in Eq.\ \eqref{cumulant}. By potentiating $\sum_j\sigma_j^x(t)$, more and more nonlocal couplings are generated. In fact, at order $k$ couplings up to distance $k+1$ are generated (see Ref.~\citen{Schmitt2018} for details). Hence, we can systematically increase the VCN by adding all distinct classical coupling terms up to a given distance $d$, which are compatible with the system’s symmetries. For example, when considering second-order corrections, the terms 
\begin{equation}
\label{H2TFIM}
\mc{H}^{(2)}_\mathrm{TFIM}=  C_1^{(2)}(t) \sum_{i=1}^N s_{i-1} s_{i}s_{i+1} s_{i+2}+ C_2^{(2)}(t) \sum_{i=1}^N s_i s_{i+3}, 
\end{equation}
that respect $\mathbb{Z}_2$ and lattice symmetries, and which expand  up to a distance $d=3$, would be added to the effective Hamiltonian, $\mc{H}_\mathrm{TFIM}=\mc{H}^{(1)}_\mathrm{TFIM}+\mc{H}^{(2)}_\mathrm{TFIM}$ (see Fig.~\ref{fig:network}).

\begin{figure*}[ht!]
	\centering
	\includegraphics[width=\textwidth]{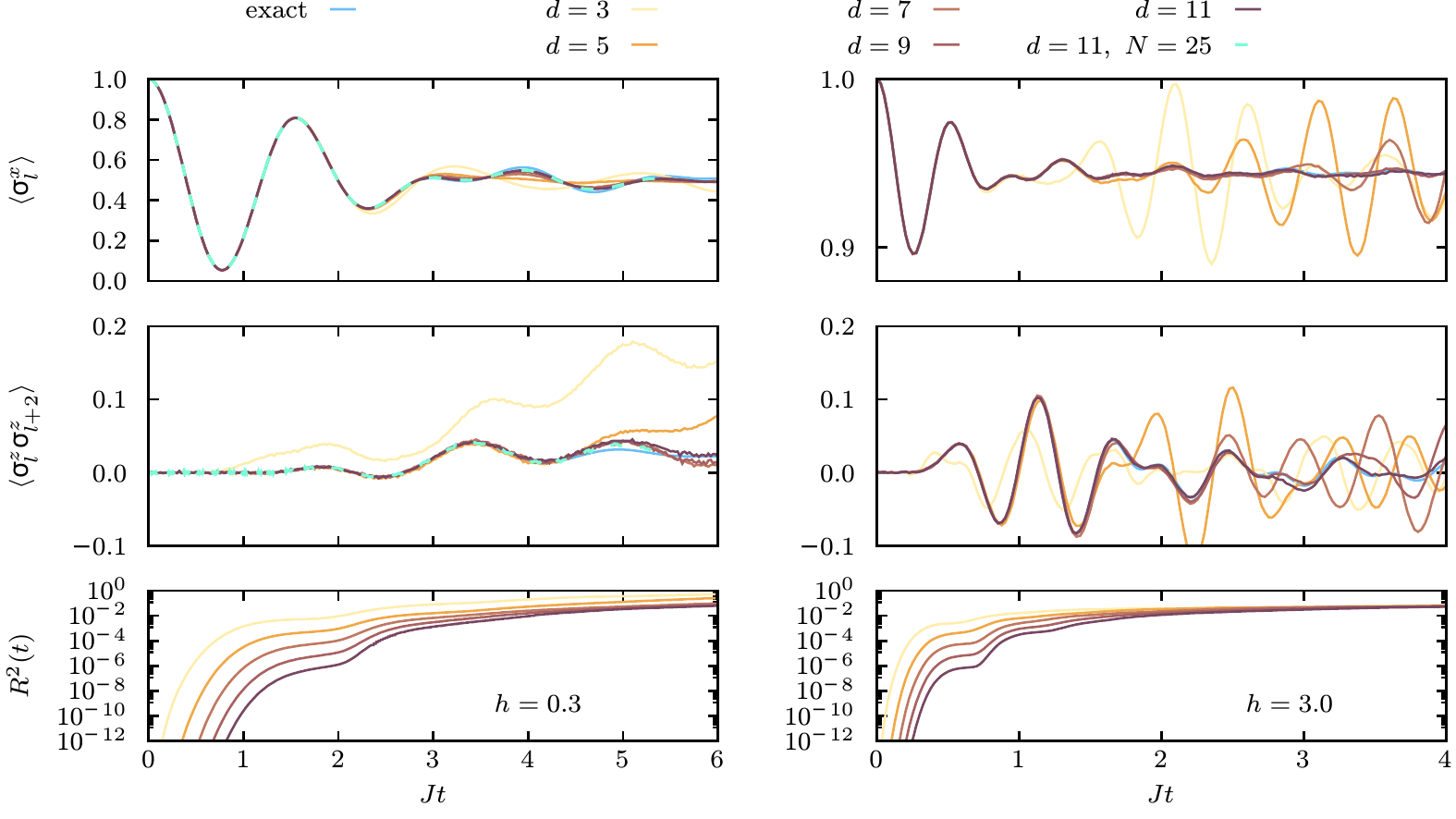}
	\caption{Dynamics of the  TFIM in quenches from $h_0=\infty$ to $h/J=0.3$ (left column) and $h/J=3$ (right column), with $N=50$. Convergence 
    of VCN solution to the exact dynamics is shown as a function of the coupling distance $d$ (see main text). Upper panels:  Transverse magnetization $\la \sigma_l^x\ra$. Middle panels: Next-to-nearest neighbor correlation function $\la \sigma_l^z \sigma_{l+2}^z \ra$. Lower panels: Integrated relative residual $R^2(t)$. Results for a different system size ($N=25$) are also shown in the quench to $h/J=0.3$ (left column) for the VCN with $d=11$.}
	\label{fig:tfim2}
\end{figure*}
Notice that the possible number of coupling terms in the classical network at $d=N$ equals the dimension of the Hilbert space of the quantum system. In the presence of translational invariance, lattice inversion symmetry, and $\mathbb Z_2$ symmetry, the symmetry allowed couplings can be obtained by generating the corresponding symmetry reduced computational basis and then identifying the domain wall configuration in each computational basis state with a coupling term in the variational wave function. In the absence of $\mathbb{Z}_2$ symmetry the computational basis configurations themselves correspond to coupling terms.

%\textcolor{red}{In the presence of translational invariance, lattice inversion symmetry, and $\mathbb Z_2$ symmetry, the symmetry allowed couplings can be obtained by generating the corresponding symmetry reduced computational basis and then identifying the domain wall configuration in each computational basis state with a coupling term in the variational wave function. In the absence of $\mathbb{Z}_2$ symmetry the computational basis configurations themselves correspond to coupling terms.}

Finally, let us point out that the classical networks presented here for the 1D TFIM can be mapped onto certain types of ANNs~\cite{Schmitt2018}.
 
\subsection{Quench protocol and results}\label{1d}

As mentioned before,  quenches in the 1D TFIM across the critical point comprise a DQPT. In that respect, an interesting class of quenches consists of going from the paramagnetic to the ferromagnetic phase. Here, we concentrate precisely on the aforementioned situation as well as  on quenches within the paramagnetic phase. In particular, we consider the initial state $\lvert\rightarrow\ra$ given in Eq.~\eqref{x-initialState}, which corresponds to the point $h_0=\infty$. Next, we compute the unitary dynamics generated by the Hamiltonian in Eq.~\eqref{Hising} with $h/J<1$ (ferromagnetic) and $h/J>1$ (paramagnetic).

We compare results for the dynamics of the TFIM obtained in three different ways: exact, pCN, and VCN. For the exact solution we exploit the integrability of the model. Via a Jordan-Wigner transformation the spin system is mapped to a model of noninteracting fermions~\cite{PFEUTY197079}, for which closed form expressions can be obtained for all quantities of interest~\cite{sachdev_2011}. The results shown are for a translationally invariant chain in the thermodynamic limit. With pCN and VCN, we consider systems with $N=50$ sites and periodic boundary conditions. On the timescales shown there is no finite-size effect in the observables. To obtain the time-evolved VCN we initialize all network couplings with zero and integrate the TDVP equation using a second-order consistent integrator with adaptive time step. Expectation values with respect to $|\psi(s)|^2$ are estimated using $8\times10^4$ samples generated by a single-spin-flip Markov chain Monte Carlo.
\begin{figure*}[tb!]
	\centering
	\includegraphics[width=\textwidth]{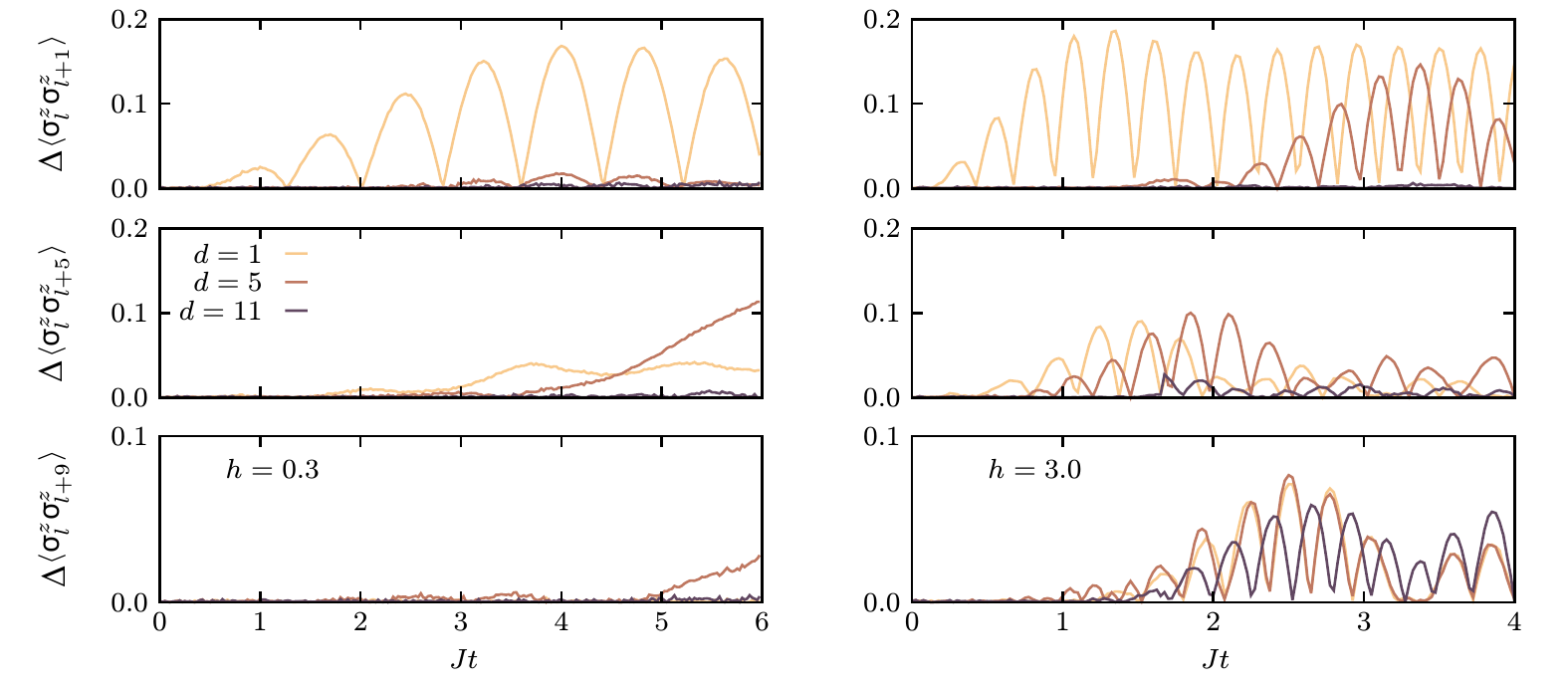}
	\caption{Accuracy of correlations in the TFIM at different distances in the quench to $h/J=0.3$ (left column) and $h/J=3$ (right column): The quantity $\Delta\la \sigma_l^z \sigma^z_{l+r} \ra$ denotes the absolute value of the difference between the TDVP result and the exact result from free fermions.}
	\label{fig:tfim_corr_accuracy}
\end{figure*}

\begin{figure*}[hbt!]
	\centering
	\includegraphics[width=\textwidth]{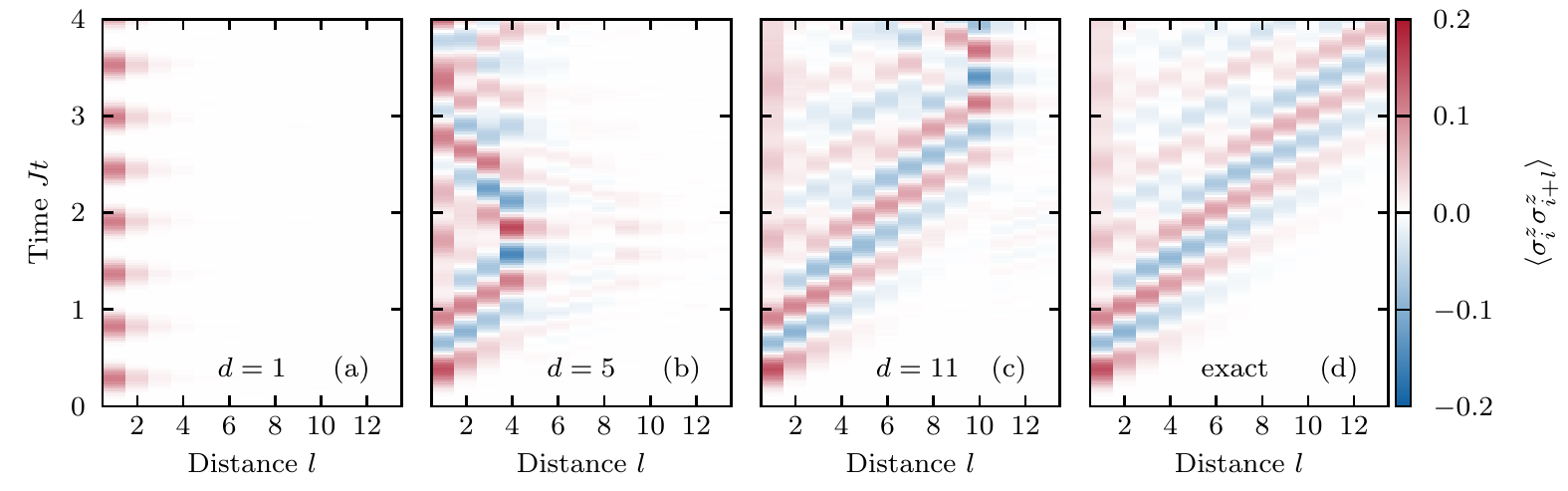}
	\caption{Correlation spreading in the Ising model for the quench to $h/J=3$, with VCNs with (a) $d=1$, (b) $d=5$, (c) $d=11$: Spreading is only captured up to the coupling distance. The exact correlation dynamics is also shown in (d).}
	\label{fig:tfim_corr}
\end{figure*}

The results of the quench dynamics are shown in Figs.~\ref{fig:tfim1}--\ref{fig:tfim_corr}. First, in Fig.~\ref{fig:tfim1}, we compare the performance of the first-order pCN given in Eq.~\eqref{H1TFIM} and its associated VCN, in a quench to $h/J=0.1$. As illustrated for the dynamics of the transverse magnetization $\la \sigma_l^x\ra$, both approaches capture very accurately the short-time behavior. However, it is the variational ansatz that yields a much more accurate description at longer times. Interestingly, when looking at the evolution of the perturbative and variational couplings, Fig.~\ref{fig:tfim1} (bottom), we can see that their dynamics start to differ approximately at the point where  discrepancies in the evolution of observables are first noted.

In Fig.~\ref{fig:tfim2}, we study the overall performance of various VCNs with different coupling distance $d$, when quenching to the ferromagnetic phase ($h/J=0.3$) and the paramagnetic one ($h/J=3$), left and right columns in Fig.~\ref{fig:tfim2}, respectively. As a principal result, we observe that the accuracy is systematically improved  upon increasing the coupling distance of the VCNs. This is not only observed from the real-time evolution of the transverse magnetization $\la\sigma^x_l \ra$, and the next-to-nearest neighbor correlation function $\la \sigma_l^z \sigma^z_{l+2}\ra$, but also from the integrated residuals $R^2(t)$, which show a systematic error convergence by  increasing $d$. Next, focusing on the transverse magnetization $\la\sigma^x_l \ra$, we see that, in both quenches, it quickly relaxes to a steady-state value: While at weak transverse field this feature can be well captured by all the considered VCNs, the situation becomes more challenging when the value of $h/J$ is large. Nevertheless, even in the latter case, the dynamics computed with the VCNs with the largest coupling distances regarded here  ($d=9, 11$) follow very closely the actual relaxation of  $\la\sigma^x_l \ra$. As for the next-to-nearest neighbor correlation function $\la \sigma_l^z \sigma^z_{l+2}\ra$, it is found that correlations at this distance are rather small in the quench to the ferromagnetic phase, whereas they are larger and oscillate between positive and negative values before decaying to zero in the quench within the paramagnetic phase. In both cases, however, it is again the highest-order VCNs that yield a better description of this correlation function, as expected. In Fig.~\ref{fig:tfim2} results corresponding to a different system size ($N=25$) are also shown for comparison in the quench to $h/J=0.3$, with $d=11$. There are no appreciable finite-size effects observed up to the accessed timescales.

The accuracy of correlations as a function of the coupling distance $d$ is analyzed in Fig.~\ref{fig:tfim_corr_accuracy}, for the two quenches considered before. In this figure we plot the deviation of the TDVP results from the exact dynamics $\Delta\la \sigma_l^z \sigma_{l+r}^z\ra$, in terms of two-point correlation functions $\la \sigma_l^z \sigma_{l+r}^z\ra$, at various distances $r$. As a general remark, we observe that the deviations from the exact result are systematically decreased by increasing the coupling distance $d$, in agreement with the results in Fig.~\ref{fig:tfim2}.  Also, it should be noted that the smaller the coupling distance $d$ is, the earlier the deviations from the exact dynamics occur, as expected from the underlying perturbatively-motivated structure of the VCNs. Moreover, we observe that in the case of large transverse field the deviations, in general, grow more than in the quench to weak fields. This is due to the fact that in the dynamics with a large transverse field, correlations develop significantly at all the considered distances in the relevant timescales, see Fig.~\ref{fig:tfim_corr} below, whereas at weak transverse field the dynamics is more local and hence correlations at large distances are rather small  (see, for instance, the correlation function on the left column of Fig.~\ref{fig:tfim2}). Lastly, note that the oscillations observed for some of the deviations arise from the fact that the variational results oscillate around the exact solution, as can be seen in Fig.~\ref{fig:tfim2}.

Finally, it is instructive to look at the correlation spreading when using VCNs with different coupling distance $d$. This is illustrated in Fig.~\ref{fig:tfim_corr} for the quench to $h/J=3$, with three VCNs with $d=1, 5, 11$, as well as the exact solution. The results shown on this figure reveal another crucial feature of VCNs: The distance for which a VCN can adequately capture the propagation of correlations is exactly determined by the coupling distance $d$. Thus, we see that the spreading of correlations can be well captured in a controlled manner by increasing the coupling distance in the structure of the VCN.  Although this result is obtained for the Ising model, we expect that it holds in general. 

\section{Quenches in the 2D $U(1)$ QLM } \label{QLM}

\subsection{Model}\label{model_QLM}
\begin{figure}[bt!]
	\centering
	\includegraphics[width=1\columnwidth]{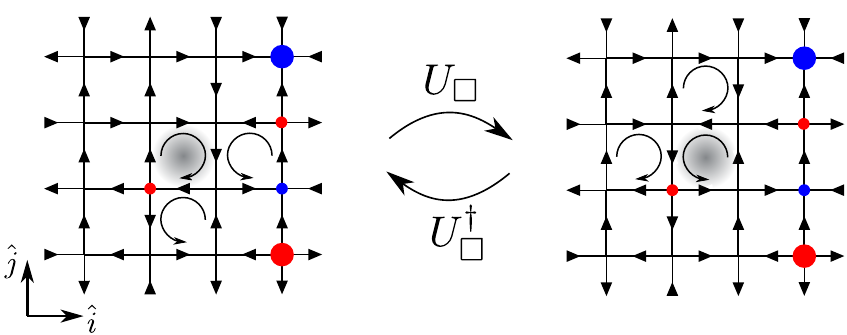}
	\caption{The 2D $U(1)$ QLM. Quantum links and electric fluxes are represented by spin-$1/2$ operators. In this figure, electric fluxes (spins) are depicted as arrows with the following  convention: Right/up (left/down) arrows correspond to $s_{x,\hat{\mu}}=+1$ ($s_{x,\hat{\mu}}=-1$), where $s_{x,\hat{\mu}}$ are the eigenvalues of $S_{x,\hat{\mu}}^{z}$. Quantum dynamics is induced by plaquette-flip operators $U_{\sq}$, $U_{\sq}^{\dg}$, where their action is schematically illustrated on the shaded plaquettes. A non-flippable plaquette is annihilated by these operators. Red (blue) circles indicate sites with a positive (negative) charge $Q_x$.   }
	\label{fig:qlm}
\end{figure}

We now show that the method presented in this paper can also be used to tackle more challenging problems such as the quantum dynamics of nonintegrable models in higher dimensions. Concretely, we study a 2D $U(1)$ quantum link model (QLM), where gauge fields are represented by spin-$1/2$ operators, defined on the links of a square lattice, and without matter degrees of freedom (see Fig.~\ref{fig:qlm}). The  Hamiltonian reads 
\begin{equation}
\label{HQLM}
H_\mathrm{QLM}=  -\mathcal{J}\sum_{\sq} (U_{\sq}+U_{\sq}^{\dg})+\lambda\sum_{\sq} (U_{\sq}+U_{\sq}^{\dg})^2 , 
\end{equation}
where the sum goes over elementary plaquettes $\sq$, and $U_{\sq} = S^+_{x, \hat{i}}S^+_{x+\hat{i}, \hat{j}} S^-_{x +\hat{j}, \hat{i}}S^-_{x, \hat{j}}$, where $S^{+/-}_{x,\hat{\mu}}$ are the standard raising/lowering spin-$1/2$ operators defined on the link that connects sites $x$ and $x+\hat{\mu}$, with the unit vectors on the 2D lattice being denoted $\hat{i}$, $\hat{j}$. For simplicity, we use spin variables normalized to 1, i.e., the eigenvalues of $S_{x,\hat{\mu}}^{z}$ are $s_{x,\hat{\mu}}=\pm 1$. This model is relevant within the context of both lattice quantum electrodynamics~\cite{Wiese2013}  and quantum spin liquids~\cite{Hermele2004}. Experimentally, there are several  proposals  for realizing lattice gauge theories in various quantum simulator settings~\cite{Wiese2013, TAGLIACOZZO2013160, Glaetzle2014,  MARCOS2014634, PhysRevX.10.021057} that could be relevant for the  2D $U(1)$ QLM studied here. In what follows, we also consider periodic boundary conditions. 

In this theory, the associated electric flux is given by the $z$ component of the spin-1/2 operator, i.e., $E_{x, \hat{\mu}}\equiv S^z_{x,\hat{\mu}}$. Thus, quantum links and electric field operators satisfy canonical commutation relations, i.e.,  $[S^z_{x,\hat{\mu}}, S^{+}_{x',\hat{\mu}'}]= S^{+}_{x,\hat{\mu}}\delta_{x,x'}\delta_{\mu,\mu'}$ and  $[S^z_{x,\hat{\mu}}, S^{-}_{x',\hat{\mu}'}]= -S^{-}_{x,\hat{\mu}}\delta_{x,x'}\delta_{\mu,\mu'}$. Moreover, the Hamiltonian~\eqref{HQLM} is invariant under local $U(1)$ transformations with generator~\cite{Wiese2013} $G(x)=\sum_{\mu}(E_{x,\hat{\mu}}-E_{x-\hat{\mu}, \hat{\mu}})$, so that $[H_{\mathrm{QLM}},G]=0$. This allows us to structure the Hilbert space into so-called superselection sectors ~\cite{Brenes2018}, with an associated background charge distribution $\{Q_{\alpha}\}$ satisfying Gauss law, i.e., $G(x)|\psi\ra=Q_x|\psi\ra$, for all $x$.  Thus, every superselection sector consists of only physical states fulfilling that the incoming and outgoing fluxes equal the charge $Q_x$ at a given site $x$, specified by the distribution $\{Q_{\alpha}\}$ defining the corresponding sector.% Further, the model in Eq.~\eqref{HQLM} has a set of global conserved quantities $\varphi_x=$

%As opposed to 1D lattice gauge theories, the dynamics of the gauge fields in 2D theories can be generated by local magnetic interactions, even when no matter fields are present~\cite{Huang2019}. That is the case considered here; correspondingly, the Hamiltonian~\eqref{HQLM} does not contain matter fields. Also, periodic boundary conditions are used in the following. 

As illustrated in Fig.~\ref{fig:qlm}, the dynamics in the QLM is generated by the first term on the right-hand side of Eq.~\eqref{HQLM}, where the plaquette operators $U_{\sq}$, $U_{\sq}^{\dg}$ induce tunneling processes between configurations with \emph{flippable} plaquettes, in which the electric fluxes form a loop with clockwise or anticlockwise orientation (a non-flippable plaquette is annihilated by such operators). The second contribution on the right-hand side of Eq.~\eqref{HQLM}, acts as a potential energy term that favors those configurations with a larger number of flippable plaquettes for $\lambda <0$. When $\lambda=-\infty$, the Hamiltonian in Eq.~\eqref{HQLM} has two $\mathbb{Z}_2$-symmetry-broken ground states, with the electric fluxes arranged such that all the plaquettes in the system are flippable; i.e., the four spins in any plaquette form a closed loop with either clockwise or anticlockwise orientation. These states, together with all other states kinetically connected to them, define a sector that we shall refer to as the \textit{fully flippable} (FF) sector.  We note that the FF sector is a subset of the superselection sector with zero charge: $G(x) |\psi\ra=0$ $\forall$ $x$.

The out-of-equilibrium dynamics of the model studied in this section has recently gained increased attention. For instance, it has been found that DQPTs occur during the unitary dynamics that follows a sudden perturbation in the Hamiltonian in Eq.~\eqref{HQLM}, when starting from one of the ground states in the FF sector~\cite{Huang2019}.  Further, in a recent paper~\cite{PhysRevLett.126.130401}, we have also shown that the model under consideration may undergo a localization-to-ergodic transition, constituting an example of a genuinely interacting 2D theory featuring disorder-free localization~\cite{Smith2017, smithPRL119, Brenes2018, PhysRevB.102.165132, PhysRevB.102.224303}. Indeed, it is found that quenches from the initial state $\lvert\rightarrow\ra$ to a QLM with $\mc{J}/\lambda < 0$ give rise to localized behavior,  such as a significant suppression of transport and a limited spread of correlations, whose origin is linked to a \emph{fragmentation}~\cite{ PhysRevX.10.011047, PhysRevB.101.174204, PhysRevB.101.125126, PhysRevLett.124.207602} of the Hilbert space into kinetically disconnected regions due to hard local constraints imposed by gauge invariance. Let us remark that in the present context, the initial state $\lvert\rightarrow\ra$ in Eq.~\eqref{x-initialState} corresponds to an equally weighted superposition of all electric flux (spin) configurations. On the other hand, the dynamics within the FF sector, for example, displays ergodic behavior, characterized by a propagation of correlations and transport quantities throughout the whole system. Subsequently, we shall consider precisely the two scenarios mentioned above. However, let us first introduce the VCNs for the QLM.
%Indeed, depending upon the quench under consideration, the system may exhibit features of ergodic or localized dynamics

\subsection{VCNs for the 2D QLM}\label{CN_QLM}

\begin{figure}[tb!]
	\centering
	\includegraphics[width=0.65\columnwidth]{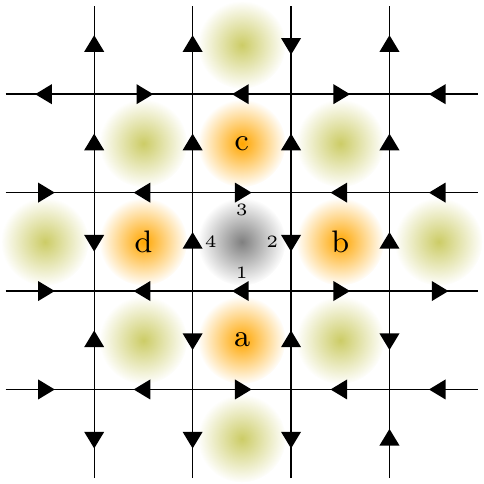}
	\caption{Convention for the definition of the operators in Eqs.~\eqref{proj_a}--\eqref{proj_d}. The projectors in these equations are defined on the bonds of the neighboring plaquettes (orange), except for those that are shared with the reference plaquette (gray). Labeling of plaquettes ($a, b, c, d$) and of spins within a plaquette ($1,2,3,4$), is done  according to their relative position as shown in this figure. Plaquettes in orange are relevant for first-order cumulants, whereas second-order cumulants involve also the plaquettes in green.}
	\label{fig:qlm_conv}
\end{figure}

Using the general scheme introduced in Sec.~\ref{CNs}, we derive pCNs and the corresponding VCNs for the QLM. The classical limit of the theory under consideration is $H_0^\mathrm{QLM}=\lambda \sum_{\sq}(U_{\sq}+U_{\sq}^{\dg})^2$. Quantum fluctuations are then induced by the term $\gamma V^\mathrm{QLM}=- \mathcal{J} \sum_{\sq}(U_{\sq}+U_{\sq}^{\dg})$, where we recognize $\gamma=-\mc{J}$. Next, we solve the equation of motion for the operator $U_{\sq}$, that is, $-i\frac{\mathrm{d}}{\mathrm{d}t}U_{\sq}(t)=[H_0, U_{\sq}]$ (and likewise for $U_{\sq}^{\dg}$). We get,
\begin{equation}
\label{U_sq}
U_{\sq}(t)=\mathrm{e}^{i\lambda \Omega_{\sq}t} U_{\sq},  \hspace{0.4cm} U_{\sq}^{\dg}(t)=\mathrm{e}^{-i\lambda \Omega_{\sq}t} U_{\sq}^{\dg},
\end{equation}
where the operator $\Omega_{\sq}$ commutes with both $U_{\sq}$ and $U_{\sq}^{\dg}$, and is given by
\begin{equation}
\label{omega}
\Omega_{\sq}= \sum_{p \in \mc{P}_{\sq}} (-\mc{A}_p+\mc{B}_p),
\end{equation}
where $\mc{P}_{\sq}=\{a,b,c,d\}$ denotes the set of neighboring plaquettes around a given plaquette (see Fig.~\ref{fig:qlm_conv}), and the operators $\mc{A}_p$ and $\mc{B}_p$ are given by
\begin{align}
\label{proj_a}
\mc{A}_a&=P_{a,1}^{\uparrow}P_{a,2}^{\uparrow}P_{a,4}^{\downarrow}, \hspace{0.3cm} 
\mc{B}_a=P_{a,1}^{\downarrow}P_{a,2}^{\downarrow}P_{a,4}^{\uparrow}, \\
\label{proj_b}
\mc{A}_b&=P_{b,1}^{\uparrow}P_{b,2}^{\uparrow}P_{b,3}^{\downarrow}, \hspace{0.3cm} \mc{B}_b=P_{b,1}^{\downarrow}P_{b,2}^{\downarrow}P_{b,3}^{\uparrow}, \\
\label{proj_c}
\mc{A}_c&=P_{c,2}^{\uparrow}P_{c,3}^{\downarrow}P_{c,4}^{\downarrow}, \hspace{0.3cm} \mc{B}_c=P_{c,2}^{\downarrow}P_{c,3}^{\uparrow}P_{c,4}^{\uparrow}, \\
\label{proj_d}
\mc{A}_d&=P_{d,1}^{\uparrow}P_{d,3}^{\downarrow}P_{d,4}^{\downarrow}, \hspace{0.3cm} \mc{B}_d=P_{d,1}^{\downarrow}P_{d,3}^{\uparrow}P_{d,4}^{\uparrow}.
\end{align}
In these definitions, the operators $P^{\uparrow, \downarrow}_{p, i}$ are projectors onto one of the components of the spins defined on the bonds of the four neighboring plaquettes; e.g.,  $P^{\uparrow}_{a, 2}= S^{+}_2S^{-}_2=(1+S^z_2)/2$, projects onto the $s_2=+1$ component of the spin living on the second link of plaquette $a$, which lies underneath the reference plaquette as in Fig.~\ref{fig:qlm_conv}. Note that the very same spin of this example is also the fourth one in the plaquette on the right side of $a$. The convention for enumerating the spins in a given plaquette and labeling neighboring plaquettes is shown in Fig.~\ref{fig:qlm_conv}. 
\subsubsection{First-order ansatz}\label{1st_qlm}

Using $\lambda$ as the unit of energy, the first-order correction of the cumulant expansion, Eq.~\eqref{H1}, gives  
\begin{equation}
\label{H1-qlm}
\mathcal{H}^{(1)}_{\mathrm{pCN-QLM}}=  i  \frac{\mc{J}}{\lambda}\sum_{\square} \Bigg( \frac{\psi_0(s_{\square})}{\psi_0(s)}  \mc{F}_{\sq}(s)^2 
  \int_0^t \mathrm{d}t'  e^{i \lambda \omega_{\square}(s)t'} \Bigg),
\end{equation}
where $\omega_{\square}(s) :=  \mc{F}_{\sq}(s)  \Omega_{\square}(s)$, and $\mc{F}_{\sq}:= U_{\square}  U_{\square}^{\dagger} -  U_{\square}^{\dagger} U_{\square}$, is a diagonal operator with possible matrix elements $\mc{F}_{\sq}(s)=+1, -1, 0$, when the reference plaquette (indicated by the subscript $\sq$) is flippable and has an anticlockwise orientation  $(+1)$, a clockwise orientation $(-1)$, or it is not flippable $(0)$. Also, we use the notation $|s_{\sq}\ra\equiv(U_{\sq}+U_{\sq}^\dg)|s\ra$, i.e., $s_{\sq}$ differs from $s$ by the flipping of a single plaquette.
$\Omega_{\square}(s)$ denotes the diagonal entries of the operator introduced in Eq.\ \eqref{omega}.

Following the prescription given by Eqs.~\eqref{integral} and \eqref{HVCN}, we can introduce a variational parameter for each one of the nine possible values that the integer variable $\omega_{\sq}(s) \in [-4,4]$ can take. Thus, using the short-hand notation

\begin{equation}
\label{eta}
\eta_{\omega_{\sq}(s)}^{(1)}(t):=\sum_{\omega=-4}^{4} \delta_{\omega_{\sq}(s), \omega} \eta_{\omega}^{(1)}(t),
\end{equation}
we can write down the corresponding variational effective Hamiltonian as
\begin{equation}
\label{H1-vqlm}
\tilde{\mc{H}}^{(1)}_{\mathrm{QLM}}= i  \frac{\mc{J}}{\lambda}\sum_{\square} \Bigg( \frac{\psi_0(s_{\square})}{\psi_0(s)}  \times (\mc{F}_{\sq}(s))^2\eta_{\omega_{\sq}(s)}^{(1)}(t) \Bigg),
\end{equation}
which contains up to nine independent (assuming translational invariance) ``first-order'' variational parameters $\eta_{\omega}^{(1)}(t)$. On the other hand, the ``zeroth-order'' contribution reads
\begin{equation}
\label{H0qlm}
 \tilde{\mc{H}}^{(0)}_{\mathrm{QLM}}=\ln(\psi_0(s))  -i E^{\ast}_{s} \eta^{(0)}(t),
\end{equation}
with a single variational parameter $\eta^{(0)}(t)$ and $E^{\ast}_{s}\equiv E_{s}/\lambda$ being dimensionless. Thus, a VCN built upon the zeroth- and first-order cumulants reads 
\begin{equation}
\label{Heff-qlm1}
\mc{H}^{(1)}_{\mathrm{QLM}}=\tilde{\mc{H}}^{(0)}_{\mathrm{QLM}}+\tilde{\mc{H}}^{(1)}_{\mathrm{QLM}}.
 \end{equation}

The first-order VCN defined above contains a  maximum coupling distance (along either the $\hat{i}$- or $\hat{j}$- direction) between plaquettes equal to 2. Indeed, for a given plaquette $\sq$, the function $\omega_{\sq}(s)$ involves only the nearest neighboring plaquettes that are shown in Fig.~\ref{fig:qlm_conv}. Thus, when plugged into the expression of the classical network, this gives rise to terms where the plaquettes that are separated the most are, for example, $b$ and $d$ in Fig.~\ref{fig:qlm_conv}.  In terms of parallel spins [e.g., spins on links $(x,x+\hat{j})$ and $(x+l \cdot \hat{i}, x+ \hat{j})$, with $l$ an integer], the maximal coupling distance is equal to 3. %Thus, the flipping of $\sq$ may affect the flippability of its nearest neighbors, which in turn may affect the flippability of the next-to-nearest neighbors of $\sq$. Hence, the maximal coupling covered by this VCN occurs between the two next-to-nearest neighbors diametrically opposed to each other  with respect to $\sq$. 

Finally, let us note  that the ansatz~\eqref{Heff-qlm1} can be explicitly recast as a classical Ising-like spin model with multiple (up to 16) spin interaction terms. Indeed, this can be achieved by rewriting the constraints in Eq.~\eqref{eta} in terms of the projectors $P_{p,i}^{\uparrow, \downarrow}$. For instance, for $\omega= 4$, we have

 %\vspace{0.1cm}
 \begin{widetext}
 
\begin{equation}
\label{non-conv}
    \delta_{\omega_{\sq}(s), \omega=4} = P_{\sq,1}^{\downarrow}P_{\sq,2}^{\downarrow}P_{\sq,3}^{\uparrow}P_{\sq,4}^{\uparrow}P_{a,1}^{\uparrow}P_{a,2}^{\uparrow}P_{a,4}^{\downarrow}P_{b,1}^{\uparrow}P_{b,2}^{\uparrow}P_{b,3}^{\downarrow}P_{c,2}^{\uparrow}P_{c,3}^{\downarrow}P_{c,4}^{\downarrow}P_{d,1}^{\uparrow}P_{d,3}^{\downarrow}P_{d,4}^{\downarrow}. %+ \mathrm{h.c.}
\end{equation}

\subsubsection{Second-order ansatz}\label{2nd_qlm}
For most of the numerical results we shall present next, it becomes important to make use of a second-order VCN. Thus, let us show explicitly how to construct such ansatz.  The second-order cumulant contains two terms as indicated in Eq.~\eqref{cumulant2}. Using the solutions in Eq.~\eqref{U_sq}, we have

	\begin{equation}
\label{H2_qlm_1}
\frac{\langle s | V(t') V(t'')|\psi_0 \rangle}{\langle s | \psi_0 \rangle}= \sum_{\square', \square''} \frac{\psi_0(s_{\square',\square''})}{   \psi_0(s)} 
 (\mc{F}_{\sq'}(s))^2(\mc{F}_{\sq''}(s_{\sq'}))^2 e^{i \lambda \omega_{\square'}(s)t'}  
e^{i \lambda \omega_{\square''}(s_{\square'})t''},
\end{equation}
\begin{equation}
\label{H2_qlm_2}
\frac{\langle s  | V(t') | \psi_0 \rangle \langle s |  V(t'')|\psi_0 \rangle}{\langle s | \psi_0 \rangle^2}
=\sum_{\square', \square''}   \frac{ \psi_0(s_{\square'}) \psi_0(s _{\square''}) }{\psi_0(s)^2} (\mc{F}_{\sq'}(s))^2(\mc{F}_{\sq''}(s))^2  e^{i \lambda \omega_{\square'}(s)t'}  
e^{i \lambda \omega_{\square''}(s)t''} ,
\end{equation}

\end{widetext}

\noindent where $|s_{\square',\square''}\ra\equiv(U_{\sq'}+U_{\sq'}^{\dg})(U_{\sq''}+U_{\sq''}^{\dg})|s\ra$. These equations then give the needed corrections to form a second-order pCN, and will be the basis to build the corresponding second-order VCN. Let us note at this point the following aspect concerning the locality of the second-order cumulant. Each of the two contributions, Eqs.~\eqref{H2_qlm_1} and \eqref{H2_qlm_2}, might, in principle, give rise to couplings at all distances. However, when we subtract them to form the overall second-order correction (see Eq.~\eqref{cumulant2}), most of the resulting terms cancel out, leaving only couplings up to some (local) coupling distance $d$. For instance, if we consider the initial state $\lvert\rightarrow\ra$  in Eq.~\eqref{x-initialState}, the ratios of initial amplitudes in the equations above reduce to 1. Then, one can easily verify that the only nonvanishing  contributions arise from overlapping plaquettes, i.e., plaquettes sharing one common link (gray and orange plaquettes in Fig.~\ref{fig:qlm_conv} ) or from plaquettes connected by a common neighboring plaquette (gray and green plaquettes in Fig.~\ref{fig:qlm_conv}). In effect, for plaquettes separated by more than one intermediate plaquette (i.e., plaquettes outside the colored region in Fig.~\ref{fig:qlm_conv}), we have that $\mc{F}_{\sq''}(s_{\sq'}) \equiv \mc{F}_{\sq''}(s)$ and $\omega_{\square''}(s_{\square'})\equiv \omega_{\square''}(s)$, and hence,  Eqs.~\eqref{H2_qlm_1} and \eqref{H2_qlm_2}, become identical.  In this case, the coupling distance (along either the $\hat{i}$- or $\hat{j}$- direction) is $d=4$ for plaquettes, and $d=5$ for parallel spins. We note that this remark also holds for other initial states that can be written as a product of local terms, as long as supports of sufficiently separated local terms do not overlap with each other, as is the case for all initial states considered in this work.

Now let us convert the second-order pCN into a second-order VCN. In the same spirit as for the first-order ansatz, the  integrals involved in the second-order correction can be written as 

\begin{widetext}
\begin{equation}
\label{integral2}
\int_0^t \mathrm{d}t' \int_0^{t'} \mathrm{d}t'' e^{i \lambda \omega_{\square'}(s)t'}  
e^{i \lambda \omega_{\square''}(s_{\square'})t''}=\sum_{\omega_1, \omega_2=-4}^{4} \delta_{\omega_{\sq'}(s), \omega_1} \delta_{\omega_{\sq''}(s_{\sq'}), \omega_2}\int_0^t \mathrm{d}t' \int_0^{t'} \mathrm{d}t'' e^{i \lambda \omega_1 t'} e^{i \lambda \omega_2 t''},
\end{equation}
for Eq.~\eqref{H2_qlm_1}, and likewise for Eq.~\eqref{H2_qlm_2}. In this case, there are 81 possible combinations of the tuple $(\omega_1,\omega_2)$; hence, we introduce 81 second-order variational parameters $\eta^{(2)}_{\omega_1,\omega_2}(t)$. Thus, using the short-hand notation
\begin{align}
\label{eta2}
\eta^{(2)}_{\omega_{\sq'}(s),\omega_{\sq''}(s_{\sq'})}(t):=\sum_{\omega_1, \omega_2=-4}^{4} \delta_{\omega_{\sq'}(s), \omega_1} \delta_{\omega_{\sq''}(s_{\sq'}), \omega_2} \eta^{(2)}_{\omega_1,\omega_2}(t),
\end{align}
and upon integrating Eq.~\eqref{H2_qlm_1}, one can make the substitution
\begin{equation}
\int_0^t dt' \int_0^{t'} dt'' \frac{\langle s | V(t') V(t'')|\psi_0 \rangle}{\langle s | \psi_0 \rangle}
\longleftarrow \sum_{\sq', \sq''} \frac{\psi_0(s_{\sq',\sq''})}{   \psi_0(s)} 
\mc{F}_{\sq'}(s)^2\mc{F}_{\sq''}(s_{\sq'})^2 \eta^{(2)}_{\omega_{\sq'}(s) \omega_{\sq''}(s_{\sq'})}(t).
\end{equation}	
Analogously, the corresponding substitution of Eq.~\eqref{H2_qlm_2}, is
\begin{equation}
\int_0^t dt' \int_0^{t'} dt'' \frac{\langle s  | V(t') | \psi_0 \rangle \langle s |  V(t'')|\psi_0 \rangle}{\langle s | \psi_0 \rangle^2}
\longleftarrow \sum_{\square', \square''}   \frac{ \psi_0(s_{\square'}) \psi_0(s _{\square''}) }{\psi_0(s)^2} \mc{F}_{\sq'}(s)^2\mc{F}_{\sq''}(s)^2 \tilde{\eta}^{(2)}_{\omega_{\square'}(s), \omega_{\square''}(s)}(t),
\end{equation}
with $\tilde{\eta}^{(2)}_{\omega_{\square'}(s), \omega_{\square''}(s)}(t)$ defining a further 81 variational parameters. Combining the previous two equations, we obtain the full, variational second-order correction: 
\begin{equation}
	\label{H2-vqlm}
	\tilde{\mc{H}}^{(2)}_{\mathrm{QLM}}=-\frac{\mc{J}^2}{\lambda^2}\tilde{\sum}_{\sq',\sq''}\Bigg( \frac{\psi_0(s_{\square',\square''})}{   \psi_0(s)}\eta^{(2)}_{\omega_{\sq'}(s),\omega_{\sq''}(s_{\sq'})}(t) \Bigg)+\frac{\mc{J}^2}{\lambda^2}\dbtilde{\sum}_{\sq',\sq''}\Bigg( \frac{ \psi_0(s_{\square'}) \psi_0( s_{\square''}) }{\psi_0(s)^2}  \tilde{\eta}^{(2)}_{\omega_{\square'}(s), \omega_{\square''}(s)}(t)  \Bigg),
\end{equation}
where $\tilde{\sum}_{\sq',\sq''}:=\sum_{\sq',\sq''}(\mc{F}_{\sq'}(s))^2(\mc{F}_{\sq''}(s_{\sq'}))^2$ and  $\dbtilde{\sum}_{\sq',\sq''}:=\sum_{\sq',\sq''}(\mc{F}_{\sq'}(s))^2(\mc{F}_{\sq''}(s))^2$. Thus, a VCN including up to second-order cumulants reads
\begin{equation}
\label{Heff-qlm2}
\mc{H}^{(2)}_{\mathrm{QLM}}=\tilde{\mc{H}}^{(0)}_{\mathrm{QLM}}+\tilde{\mc{H}}^{(1)}_{\mathrm{QLM}}+	\tilde{\mc{H}}^{(2)}_{\mathrm{QLM}},
\end{equation}
with the expressions given in Eqs.~\eqref{H0qlm}, \eqref{H1-vqlm}, and \eqref{H2-vqlm}.

Finally, let us point out that we restrict the resulting VCNs to have the same locality as the underlying pCNs;
i.e., the sums over plaquette $\square''$ in \eqref{H2-vqlm} are restricted to the nearest or next-nearest neighbors of plaquette $\square'$ (see Fig. \ref{fig:qlm_conv}).
\end{widetext}

\subsection{Quench protocol and results}\label{quenchQLM}

\begin{figure*}[tb!]
	\centering
	\includegraphics[width=1\textwidth]{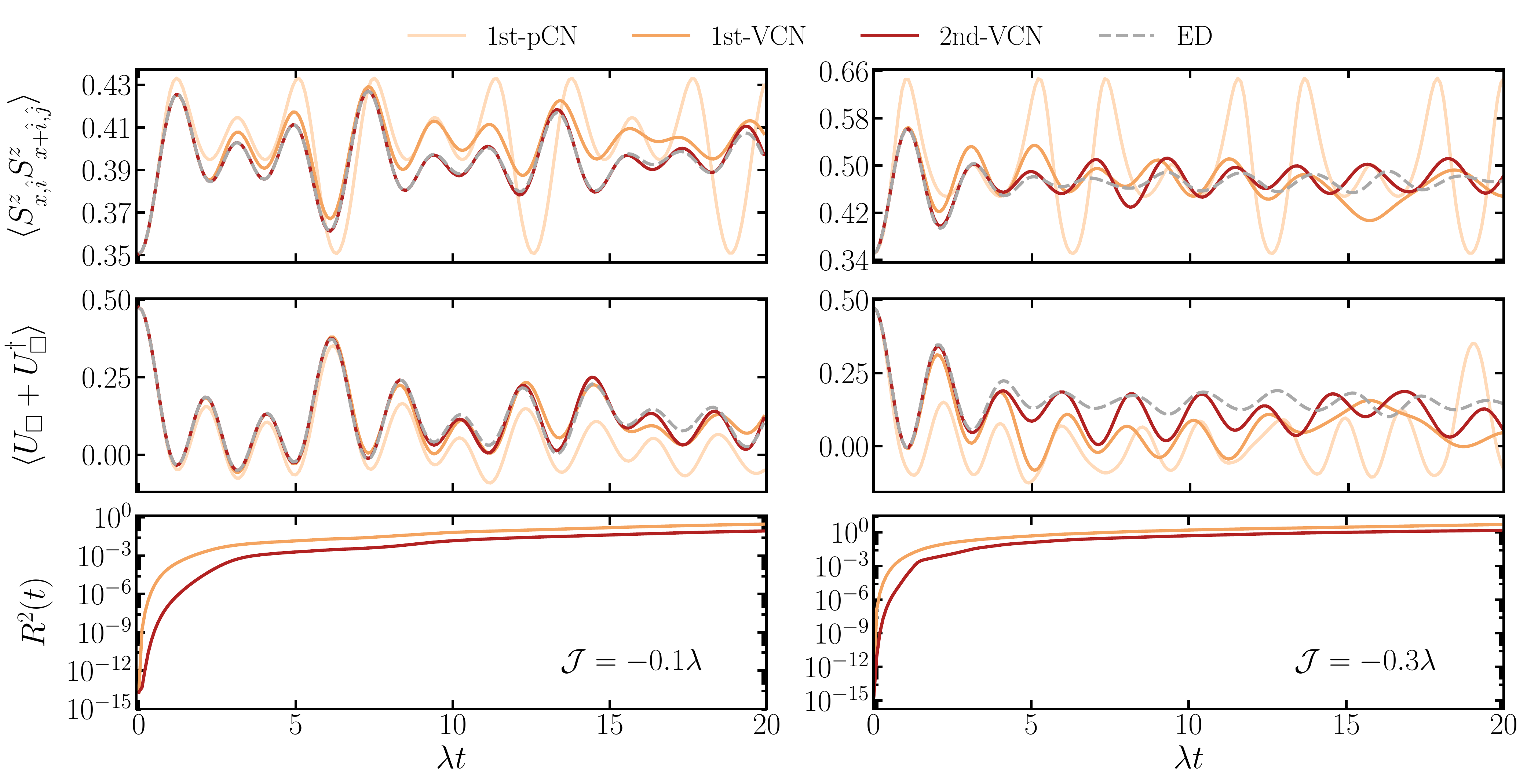}
	\caption{Comparison of  perturbative, variational, and exact dynamics in a quasi-1D  ladder of $2 \times 10$ plaquettes (40 spins): Left and right columns show the quench dynamics from $\lvert\rightarrow\ra_\mathrm{FF}$ to $\mathcal{J}/\lambda=-0.1$ and $\mathcal{J}/\lambda=-0.3$, respectively.   Upper row: Dynamics of the nearest-neighbor spin-spin correlator $\la S^z_{x, \hat{i}}S^z_{x+\hat{i}, \hat{j}} \ra $. Middle row: Dynamics of the (dimensionless) mean kinetic energy per plaquette $\la U_{\sq}+U_{\sq}^{\dg}\ra$. Lower row: Integrated residuals $R^2(t)$, of the first- and second-order VCNs.}
	\label{fig:results1-qlm}
\end{figure*}

In order to test the performance of the VCNs defined in Eqs.~\eqref{Heff-qlm1} and~\eqref{Heff-qlm2}, we investigate several quantum quenches in the 2D QLM. First, we benchmark our method by considering a setting where it is still possible to perform exact diagonalization (ED) for relatively large system sizes. This is achieved by regarding a quasi-1D ladder with dynamics restricted to a given superselection sector. In this type of setting, we study quenches from both uniform and nonuniform initial states. Next, we consider situations well beyond the scope of ED. Namely, we regard a truly two-dimensional configuration, and quenches from the initial state in Eq.~\eqref{x-initialState}, which involves all the superselection sectors. As explained below, for the particular physical scenario considered in this case (disorder-free localization), our method is able to yield sufficiently accurate results for long timescales that could hardly be accessed with any other state-of-the-art numerical technique. Unless otherwise stated, in all the examples discussed in this section, the TDVP equations are solved using a 4th-order Runge-Kutta integrator with step size $\Delta t= 0.1\lambda^{-1}$.

\subsubsection{Benchmark in quasi-1D ladders: uniform initial states}\label{benchmarkQLM}

As a first benchmark,  we consider a quasi-1D ladder of $2\times10$ plaquettes, i.e., 40 spins, with dynamics restricted to a given superselection sector. Concretely, we consider the FF sector, consisting of all the states with zero charge that are kinetically connected to the two maximally flippable configurations, as explained in Sec.~\ref{model_QLM}. As shown below, with these settings, one can still carry out efficient ED calculations. Indeed, the key idea is that, since the Hamiltonian in Eq.~\eqref{HQLM} adopts a block-diagonal form, with each block corresponding to a given superselection sector, a state that belongs to one of such sectors will remain in that sector during the course of the dynamics generated by $H_\mathrm{QLM}$. Thus, one needs to consider only the portion of the Hilbert space that is relevant for the chosen sector.%, and therefore, exact diagonalization can be performed for system sizes as the ones mentioned above. 

 \begin{figure*}[bt!]
	\centering
	\includegraphics[width=1\textwidth]{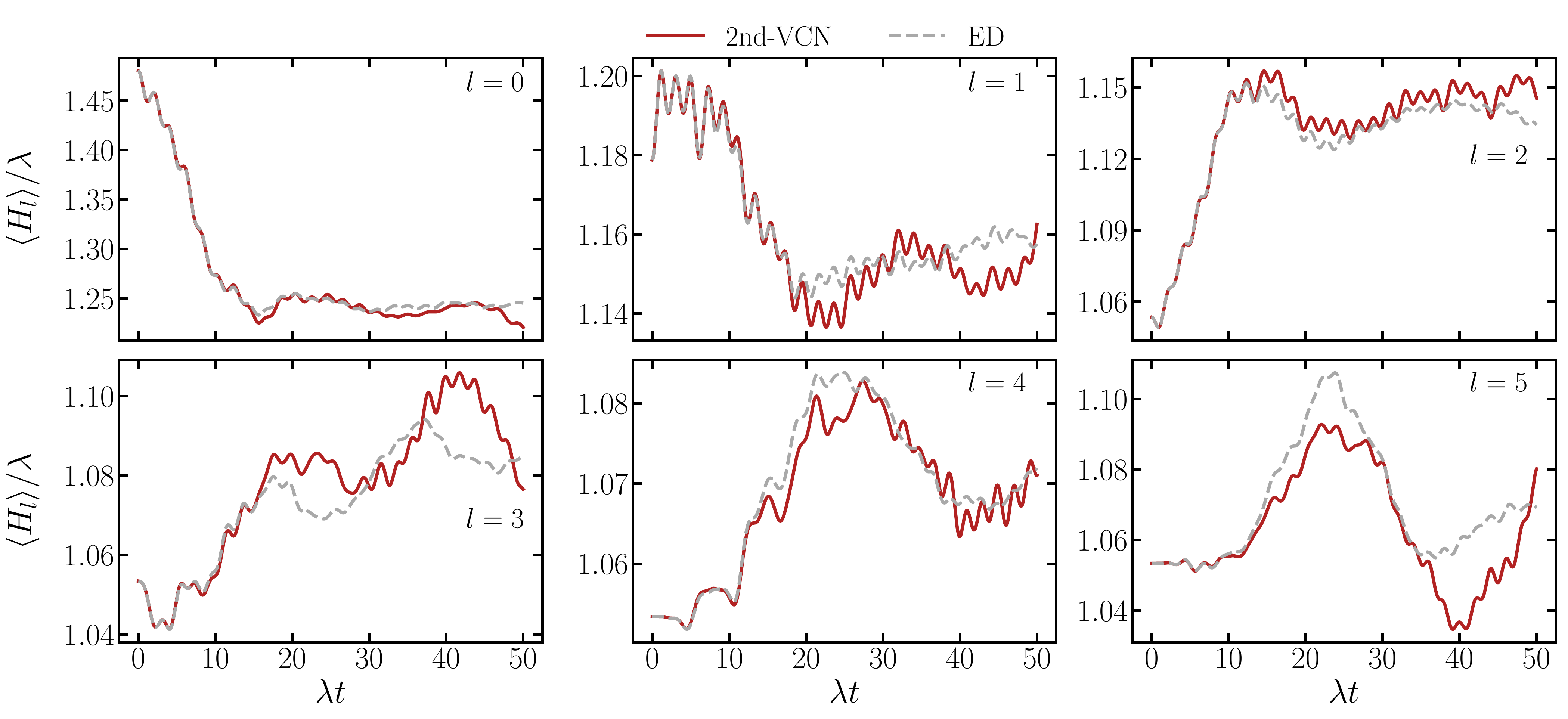}
	\caption{Energy of the $l$th column with $l=0, \dots, 5$, for the quench from the nonuniform initial state $\mathscr{P}\lvert\rightarrow\ra_\mathrm{FF}$, containing an excess of energy around $l=0$ (see main text) to $\mc{J}/\lambda=-0.1$. We compare the TDVP solution using a second-order VCN [Eq.~\eqref{Heff-qlm2}] with ED results.}
	\label{fig:results4-qlm}
\end{figure*}

The quench protocol considered here is the following: We initialize the system in an equally weighted superposition of all the basis states spanning the FF sector. We denote such a superposition  as $\lvert\rightarrow\ra_\mathrm{FF}$. Next, we evolve the system with the Hamiltonian in Eq.~\eqref{HQLM}, with a finite value of $\mc{J}/\lambda$. Later, we will study similar quenches but starting from nonuniform initial states.  Note that for a   $2\times10$ system, the dimension of the FF sector is 17906. Therefore, in the TDVP calculations, we can also carry out an exact enumeration of states, i.e., we can explicitly sum over all relevant spin configurations in expressions such as the expectation value in Eq.~\eqref{obs}, rather than performing a Monte Carlo sampling.

Let us emphasize that the dynamics in the FF sector is interesting because significant correlations may develop in the entire spatial extent of the system, as we have recently shown~\cite{PhysRevLett.126.130401}. Therefore, this constitutes a highly challenging scenario for our method, since a VCN can adequately capture the buildup of correlations up to a distance specified, essentially,  by the order of the cumulant expansion, as we have already argued. 

 The results of this benchmark are shown in Fig.~\ref{fig:results1-qlm}, for quenches to $\mc{J}/\lambda=-0.1$ and  $\mc{J}/\lambda=-0.3$, left and right columns of Fig.~\ref{fig:results1-qlm}, respectively. We compare the dynamics computed with a first-order pCN, a first-order VCN [Eq.~\eqref{Heff-qlm1}], a second-order VCN [Eq.~\eqref{Heff-qlm2}], and via ED. Particularly, we focus on two observables: the nearest-neighbor spin-spin correlation function $\la S^z_{x,\hat{i}} S^z_{x+\hat{i}, \hat{j}}\ra$ (i.e., $\la S^z_1 S^z_2 \ra$, with the convention in Fig.~\ref{fig:qlm_conv}), and the mean kinetic energy (up to a factor of $-\mc{J}$) per plaquette $\la U_{\sq} +U^{\dagger}_{\sq}\ra$. Besides, we also show 
 the integrated relative residuals $R^2(t)$ of the first- and second-order VCNs, as a measure of their accuracy.

As expected, it is the second-order VCN that provides the most accurate results in the studied quenches. This can be seen from both the evolution of the observables and the growth of the residuals. In particular, for the quench to $\mc{J}/\lambda=-0.1$, the second-order VCN captures remarkably well correlations at short distances (left column, upper row), as well as the oscillatory behavior of off-diagonal observables (left column, middle row), in the entire range of accessed timescales $\lambda t= 20$. As anticipated, in the quench with a bigger strength of the perturbation $\mc{J}/\lambda=-0.3$, the description in terms of all of the considered wave functions, breaks down at earlier times. Yet, the second-order VCN still yields a rather excellent agreement with ED  up to significant timescales  $\lambda t  \sim 7$. In all cases, the VCNs outperform the first-order pCN regarded here. 
 
 Next, we point out that, as illustrated better in the second quench (right panel of Fig.~\ref{fig:results1-qlm}), the oscillations of both $\la S_{x,\hat{i}}^zS_{x+\hat{i},\hat{j}}^z\ra $ and $\la U_{\sq}+U^{\dg}_{\sq}\ra$ \emph{decay} to some steady-state value
 in agreement with the expectation of ergodic behavior in the FF sector~\cite{PhysRevLett.126.130401}. Importantly, such decay of the oscillations is captured by the TDVP solutions. On the contrary, the pCN fails to capture this crucial feature, although it gets right the frequency of the oscillations up to considerable timescales. 
Finally, let us  emphasize that the  results displayed in Fig.~\ref{fig:results1-qlm} corroborate once more the \emph{controlled} nature of VCNs, in that the  accuracy of a VCN can be improved systematically either by adding higher-order terms of the cumulant expansion or by decreasing the strength of  quantum fluctuations.

   \begin{figure*}[tb!]
	\centering
	\includegraphics[width=1\textwidth]{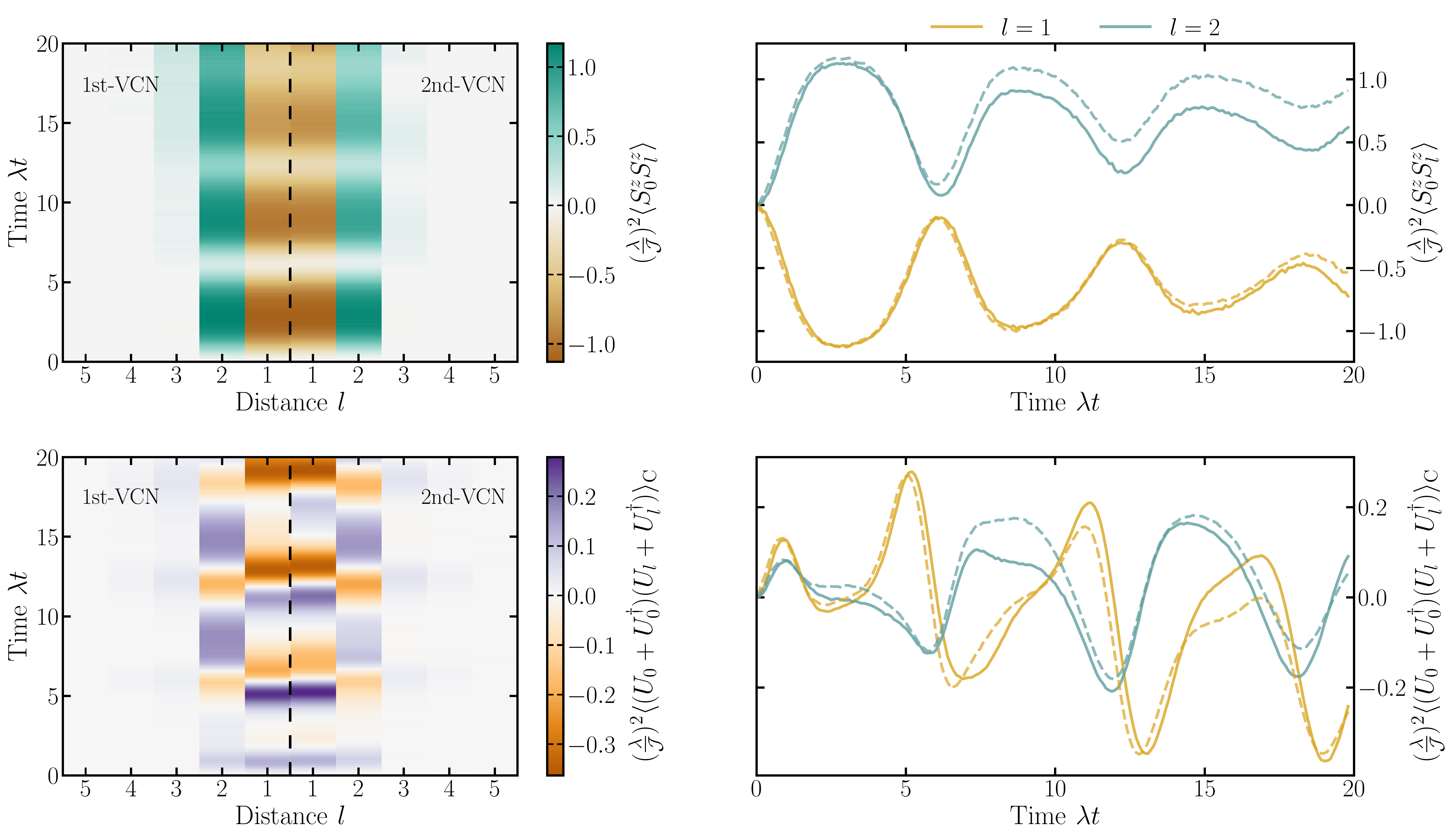}
	\caption{Correlation dynamics  in the 2D QLM for a quench from  $\lvert\rightarrow\ra$ to $\mc{J}/\lambda=-0.1$.  Left column: Spatiotemporal buildup of quantum correlations computed with  a first-order VCN (left side), and a second-order VCN (right side). Upper panel: Spin-spin correlation function $\la S^z_0 S^z_l\ra\equiv \la S^z_{x, \hat{j}} S^z_{x+l\cdot\hat{i}, \hat{j}}\ra$, between parallel  links separated by a distance $l$; lower panel: plaquette-plaquette connected correlation function $\la (U_0+U_0^{\dg})(U_l+U_l^{\dg})\ra_\mathrm{C}$ (see  main text). Right column: Individual cuts along the distances with the dominant signal, $l=1, 2$, of the corresponding quantities on the left column; second-order VCN (solid), first-order VCN (dashed). Results for a $100\times100$ system ($2\times10^4$ spins).}
	\label{fig:results2-qlm}
\end{figure*}

\subsubsection{Benchmark in quasi-1D ladders: nonuniform initial states}\label{benchmarkQLM}

Here, we consider the same setting as before, namely, a $2\times10$ quasi-1D ladder, with dynamics restricted to the FF sector. The chosen initial state, however, is nonuniform. In particular, the initial condition is created by adding a line defect with subextensive energy contribution to the state $\lvert\rightarrow\ra_\mathrm{FF}$. This is achieved by applying the operator 

\begin{equation}
    \label{defect}
   \mathscr{P} =\prod_{\sq \in  \mathscr{C}_0} 1 +(U_{\sq}+U_{\sq}^{\dg})^2,
\end{equation}
upon the state $\lvert\rightarrow\ra_\mathrm{FF}$, where $\mathscr{C}_d$ denotes the set of plaquettes in the $l$th column. Thus, $\mathscr{P}\lvert\rightarrow\ra_\mathrm{FF}$ represents a state with a line energy defect along column $l=0$. The resulting quench dynamics obtained using both a second-order VCN and ED are analyzed in terms of the evolution of the total energy of column $l$, which is given by
\begin{equation}
    \label{energy_col}
   H_l=\sum_{\sq \in  \mathscr{C}_l} - \mathcal{J} (U_{\sq}+U_{\sq}^{\dg})+\lambda (U_{\sq}+U_{\sq}^{\dg})^2.
\end{equation}
The results are shown in Fig.~\ref{fig:results4-qlm}. One can observe that the second-order VCN captures quite well, in a quantitative way, the propagation of the line defect at all distances up to a time $\lambda t \approx 20$. After this point, the TDVP solution is not exact anymore, but it still follows qualitatively the exact dynamics to the largest accessed time, $\lambda t =50$.  The quantity in Eq.~\eqref{energy_col} can serve as a probe to distinguish between ergodic and nonergodic behavior in the QLM as done in Ref.~\citen{PhysRevLett.126.130401}.

\subsubsection{Quenches in a 2D lattice}\label{results2DQLM}

After having assessed the performance of our method in quasi-1D ladders, we now consider a truly 2D setting.  Moreover, we shall regard regimes where access to ED is computationally prohibited. Concretely, the following quench protocol is carried out: The system is initialized in the state $\lvert\rightarrow\ra$ given in Eq.~\eqref{x-initialState} and then evolved with the Hamiltonian in Eq.~\eqref{HQLM} at finite $\mc{J}/\lambda$. As mentioned before, in the present context the state $\lvert\rightarrow\ra$ can be thought of as an equally weighted superposition of all the electric flux (spin) configurations, thereby involving all the superselection sectors. Thus, apart from very small system sizes, ED techniques become impractical. In the following, we show results for a system of $100 \times 100$ plaquettes ($2\times 10^4$ spins). In all the calculations considered here, a Metropolis Monte Carlo sampling was performed with $10^6$ sweeps at each time instance, and single-spin-flip updates.

As explained in our recent work~\cite{PhysRevLett.126.130401}, it is interesting to investigate quenches from the  initial state $\lvert \rightarrow\ra$ to the QLM in Eq.~\eqref{HQLM}, as this scenario gives rise to (disorder-free) many-body localized dynamics, even though both the Hamiltonian and the initial condition are homogeneous.  In effect, when quenching from the  initial state $\lvert\rightarrow\ra$, not only the transport of energy is highly suppressed, but also the spreading of correlations is drastically constrained~\cite{PhysRevLett.126.130401}. This means that relevant correlations  can only develop at very short distances. Due to this reason, we expect  low-order VCNs to capture  very well the main features of the exact quantum dynamics in the present case.

\begin{figure*}[tb!]
	\centering
	\includegraphics[width=1\textwidth]{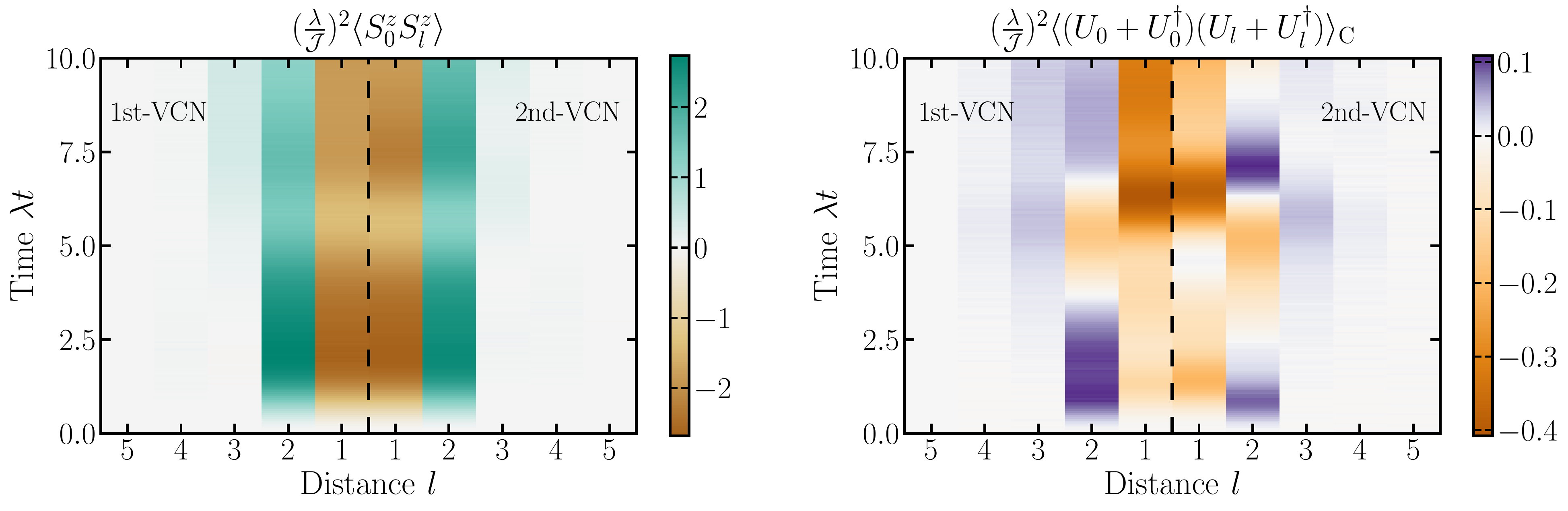}
	\caption{Correlation dynamics in the 2D QLM for a quench from  $\lvert\rightarrow\ra$ to  $\mc{J}/\lambda=-0.3$.  Left panel: Spin-spin correlation function $\la S^z_0(t) S^z_l(t)\ra \equiv \la S^z_{x, \hat{j}}(t) S^z_{x+l\cdot\hat{i}, \hat{j}}(t)\ra$, between parallel  links separated by a distance $l$. Right panel: Plaquette-plaquette connected correlation function $\la (U_0+U_0^{\dg})(U_l+U_l^{\dg})\ra_\mathrm{C}$ (see main text). Results obtained with  a first-order VCN (left side), and a second-order VCN (right side), for a $100\times100$ system ($2\times10^4$ spins). Here, a time step  $\Delta t= 0.05\lambda^{-1}$ was used to integrate the TDVP equations.}
	\label{fig:results3-qlm}
\end{figure*}

The resulting dynamics is shown in Figs.~\ref{fig:results2-qlm} and \ref{fig:results3-qlm}, where quenches to the points $\mc{J}/\lambda=-0.1$ and $\mc{J}/\lambda=-0.3$, are considered, respectively. Let us focus first on the quench to $\mc{J}/\lambda=-0.1$  (Fig.~\ref{fig:results2-qlm}). The left column shows the spatiotemporal buildup of quantum correlations in terms of two correlation functions, namely, the spin-spin correlation function $\la S^z_0(t) S^z_l(t)\ra\equiv \la S^z_{x, \hat{j}}(t) S^z_{x+l\cdot\hat{i}, \hat{j}}(t)\ra$, between parallel spins separated by a distance $l$, and the connected two-point plaquette-flip correlation function $\la (U_0+U_0^{\dg})(U_l+U_l^{\dg})\ra_\mathrm{C}\equiv\la (U_0+U_0^{\dg})(U_l+U_l^{\dg})\ra - \la (U_0+U_0^{\dg})\ra^2$.  In both cases, we compare the results obtained using a first-order VCN (left-hand side of the colormap), and a second-order VCN (right-hand side of the colormap). As a first observation, we note the rather confined spreading of correlations. As already mentioned,  this is due to the disorder-free localization mechanism that takes place in the situation considered here. We stress that this is not an artifact of the VCNs. Indeed,  a first-order VCN, see Eq.~\eqref{Heff-qlm1}, has a coupling distance $d=2$, for plaquettes, and $d=3$, for parallel spins. On the other hand, a second-order VCN, see Eq.~\eqref{Heff-qlm2}, has coupling distances $d=4, 5$, for plaquettes and parallel spins, respectively. Therefore, according to our previous analysis (see Fig.~\ref{fig:tfim_corr}, in particular), we expect that both of these VCNs would be able to  account for the spreading of correlations, at least, up to the aforementioned distances. However, we find that significant correlations are developed predominantly no further than a distance $l=3$, at least, up to the accessed timescales.

A second observation is that the results obtained with both VCNs exhibit a very similar qualitative behavior. On the right column of Fig.~\ref{fig:results2-qlm}, we make a more quantitative comparison by plotting individual cuts along the distances with the largest signal, i.e., $l=1, 2$. We notice that the oscillations in both cases have pretty much the same frequency, whereas the observed discrepancies are potentially due to the fact that the first-order ansatz cannot quite capture the decay of oscillations, as also  seen in Fig.~\ref{fig:results1-qlm} for the quasi-1D ladder.    

Finally, in Fig.~\ref{fig:results3-qlm}, we show  the dynamics of correlations for the quench to $\mc{J}/\lambda=-0.3$. We note that, apart from the fact that the dynamics gets accelerated due to the bigger strength of the off-diagonal perturbation,  remarks similar to the ones for the previous quench also hold in this case. In particular, we see that the correlation spreading is constrained, too, with significant correlations growing only in the spatial region $l \le 4$. Once more, this is a manifestation of the disorder-free localization phenomenon that occurs in the dynamics of the 2D QLM for the type of quenches studied in this section. 

\section{Summary and outlook} \label{conclusions}
We have introduced a  numerical variational scheme for the study of dynamics in correlated quantum  systems in one and higher dimensions. Our method relies on an efficient representation of the many-body wave function, in terms of complex networks of classical spin variables. This class of variational wave functions, termed VCNs, is similar to ANNs. Crucially, VCNs  can  be constructed according to a controlled prescription, as it has been underlined and explicitly shown in this paper.   
Going beyond Ref.~\citen{PhysRevLett.126.130401} where VCNs were introduced only for the particular case of the 2D QLM, here we presented VCNs as a general numerical framework that can be applied to any interacting quantum lattice model, regardless of spatial dimensionality, provided that an expansion around a well-defined classical limit is possible.
%\commentPRB{Going beyond Ref.~\citen{2020arXiv200304901K} where VCNs were introduced only for the particular case of the 2D QLM, here we presented VCNs as a general numerical framework that can be applied to any interacting quantum lattice model, regardless of spatial dimensionality, provided that an expansion around a well-defined classical limit is possible.}

%Moreover, the ideas presented in this work are of general use and can be applied, in principle, to study any interacting quantum lattice model, regardless of spatial dimensionality, provided that an expansion around a well-defined classical limit is possible.

We have illustrated the way in which our method works and its range of applicability by studying quantum quenches in two very distinct models. First, we considered  the paradigmatic 1D TFIM, which serves as an ideal testing ground for our method as it can be solved exactly~\cite{PFEUTY197079}. Moreover, the 1D TFIM presents the advantage that the resulting VCNs have a relatively simple and intuitive form, which is in fact, that of classical Ising models. Further, for this example we have shown, in a rather intuitive manner, how the perturbatively motivated architecture of VCNs can be systematically built upon by incorporating classical couplings that expand over a certain coupling distance, which is closely related to the order of the underlying cumulant expansion. Then, we characterized the performance of various VCNs in terms of the real-time evolution of  two-point correlation functions at varying distance, local off-diagonal observables, and the integrated residuals. In general, we  found that the basic architecture of VCNs allows for a systematic improvement in the accuracy of such results by adding higher-order terms according to the controlled procedure referred to above. We have verified that this statement remains true when varying system size and quenching to different points in the parameter space of the associated Hamiltonian. % (see also remarks below).

Next, we  studied the out-of-equilibrium dynamics in a genuinely interacting two-dimensional lattice gauge theory, a $U(1)$ QLM. As opposed to the Ising model, in this case the resulting VCNs are highly nonconventional spin models; see, for example, Eq.~\eqref{non-conv}. Yet,  the same general ideas apply in their construction. We have shown this by explicitly calculating first- and second-order VCNs, which were then used for computing the dynamics of the  QLM in different scenarios. On the one hand, we considered  quasi-1D ladders of $2 \times 10$ (40 spins), with dynamics restricted to the FF sector. This allowed us to use ED for the purpose of benchmarking.  We emphasize, however, that this is a rather challenging scenario for our method, as it has been found~\cite{PhysRevLett.126.130401} that the dynamics in the FF sector embodies considerable quantum correlations that propagate throughout the entire system. Yet, a remarkable quantitative agreement with ED was obtained. In particular, we found that second-order VCNs yield very accurate results for short-range correlations and local off-diagonal observables, up to long timescales. Importantly, we tested our method using not only homogeneous initial conditions but also nonuniform ones with a spatial energy inhomogeneity along a given column. Once again, we found that second-order VCNs are able to accurately describe the propagation of such a  line defect at all distances up to long timescales.  Next, we considered situations where ED and other state-of-the-art techniques become computationally inaccessible.  Namely, we studied quenches starting from an equally weighted superposition of all spin configurations, thereby involving all superselection sectors. This part of our study was done in systems with $100 \times 100$ plaquettes ($2\times 10^4$ spins).  Here, however, the crucial observation that the QLM exhibits disorder-free localization dynamics~\cite{PhysRevLett.126.130401} enabled us to employ reliably our VCNs, since such localization mechanism yields a strong suppression of correlation spreading.

Overall, the quantum quenches discussed above  have allowed us to characterize the accuracy and versatility of our methodology. As a principal observation, we note that upon adding higher-order couplings in the structure of a VCN, its accuracy can be improved in a controlled way. In addition, higher-order couplings also account for another crucial feature: correlation spreading is properly captured up to a spatial scale determined by the maximum coupling distance included in the architecture of  the VCN.  On the other hand, at a fixed order of the cumulant expansion, the accuracy can also be systematically increased by reducing the strength of the off-diagonal perturbation. As a rule of thumb, we expect a low-order VCN to give sufficiently accurate results up to a timescale set by the inverse of the perturbation strength~\cite{Schmitt2018}. However, such timescales can be further prolonged when considering higher-order VCNs. Besides, for a given order of the cumulant expansion, in general, a VCN outperforms its corresponding pCN at all relevant timescales, and in some cases, the former can account for important features of the quantum dynamics, which are just beyond the scope of the latter, such as the relaxation of observables towards a steady-state value. We have checked that all the remarks made above also hold when varying system size, and quenching to different points in the parameter space of the studied Hamiltonians. 

Naturally, our method is not exempt from limitations and drawbacks, which can be understood from the considerations made in the previous paragraph. First and foremost,  the addition of higher-order terms is accompanied by an exponential growth in complexity, so in practice, we are limited to a given order of the cumulant expansion. Moreover, we also know that a VCN will fail at capturing correlations beyond their maximum coupling distance, as illustrated very clearly in Fig.~\ref{fig:tfim_corr}. Such coupling distance is also determined by  the order of the cumulant expansion. Also, in general, we expect a breakdown in the description of the evolution of observables after a timescale set by the inverse of the perturbation strength, at least, as far as low-order VCNs are concerned.

Regarding possible further applications, there are several interesting routes that one could explore employing VCNs. A particularly promising one consists of formulating hybrid approaches that combine VCNs with ANNs, so as to mitigate some of the drawbacks sustained by both kinds of generative machines. Also, as we have already mentioned, the applicability of the approach presented in this paper is not fundamentally restricted by spatial dimensionality. Thus, in this respect, the following models arise as  natural extensions of the systems studied here: 2D TFIMs, 2D bosonic and fermionic Hubbard models, 2D QLMs in the presence of a dynamical matter field,  3D quantum spin ice models, as well as other kinetically constrained models. Lastly, regarding VCNs as generative machines  also raises the possibility of employing them for tasks other than solving the real-time dynamics of quantum many-body systems. In particular, VCNs could be used for addressing ground-state search problems, with an optimization procedure guided by a conventional variational principle that minimizes the energy functional, rather than using a TDVP. %Naturally, the feasibility of the ideas discussed in this paragraph requires further exploration. 

Finally, let us emphasize that the approach presented in this paper not only allows us to address theoretical questions regarding higher-dimensional systems but also could provide a theoretical description of recent experiments in two dimensions using quantum simulators.  Among such experiments are studies of two-dimensional many-body localization dynamics employing bosons~\cite{Choi1547} and fermions~\cite{PhysRevLett.114.083002, PhysRevX.7.041047}, experiments probing transport properties of various 2D lattice systems~\cite{Hild2014, PhysRevX.10.011042}, as well as investigations on other aspects of the  quench dynamics of several 2D quantum spin systems~\cite{Labuhn2016, Lienhard2018, Guardado-Sanchez2018}. Most of the aforementioned experiments explore regimes that are still inaccessible to simulation methods relying on classical resources, thereby calling for a larger input from theoretical and numerical sides. We believe that the method introduced in this work does constitute an important step into this direction.   
\\
 
\section*{Acknowledgments} 

We are grateful to H. Burau, J. Chalker, M. Dalmonte, R. Moessner, and G.-Y. Zhu for helpful discussions. P.K. acknowledges the support of the Alexander von Humboldt Foundation and the Ministry of Science and Higher Education of the Russian Federation  (NUST MISiS Grant No. K2-2020-038).	This project has received funding from the European Research Council (ERC) under the European Union’s Horizon 2020 research and innovation program (Grant Agreement No. 853443), and M.H. further acknowledges support by the Deutsche Forschungsgemeinschaft via the Gottfried Wilhelm Leibniz Prize program. Y.-P.H. receives funding from the European Union's Horizon 2020 research and innovation program under the Marie Sk\l odowska-Curie Grant Agreement No. 701647. M.S. was supported through the Leopoldina Fellowship Programme of the German National Academy of Sciences Leopoldina (LPDS 2018-07) with additional support from the Simons Foundation. 
 
\bibliography{BIB_1}

\end{document}